\documentclass[aps,pra,twocolumn,superscriptaddress,longbibliography,letterpaper]{revtex4-1}
\usepackage{graphicx}
\usepackage{amsmath}
\usepackage{amssymb}
\usepackage{esint}
\usepackage{verbatim}
\usepackage{xcolor}
\usepackage{soul}
\usepackage{siunitx}
\DeclareSIUnit\pixel{px}
\usepackage{lineno}
\usepackage{hyperref}
\usepackage{braket}
\usepackage[version=4]{mhchem}

\newcommand{\be}{\begin{equation}}
\newcommand{\ee}{\end{equation}}
\newcommand{\ba}{\begin{array}}
\newcommand{\ea}{\end{array}}
\newcommand{\bqa}{\begin{eqnarray}}
\newcommand{\eqa}{\end{eqnarray}}

\begin{document}

\title{Observing polarization patterns in the collective motion of nanomechanical arrays}

\author{Juliane Doster}
\thanks{These two authors contributed equally to this work}
\affiliation{University of Konstanz, Department of Physics, Universitätsstr. 10, 78457 Konstanz, Germany}
\author{Tirth Shah}
\thanks{These two authors contributed equally to this work}
\affiliation{Max Planck Institute for the Science of Light, Staudtstr. 2, D-91058 Erlangen, Germany}
\affiliation{Friedrich-Alexander University Erlangen-Nürnberg (FAU), Department of Physics, Staudtstr. 7, D-91058 Erlangen, Germany}
\author{Thomas Fösel}
\affiliation{Max Planck Institute for the Science of Light, Staudtstr. 2, D-91058 Erlangen, Germany}
\affiliation{Friedrich-Alexander University Erlangen-Nürnberg (FAU), Department of Physics, Staudtstr. 7, D-91058 Erlangen, Germany}
\author{Florian Marquardt}
\affiliation{Max Planck Institute for the Science of Light, Staudtstr. 2, D-91058 Erlangen, Germany}
\affiliation{Friedrich-Alexander University Erlangen-Nürnberg (FAU), Department of Physics, Staudtstr. 7, D-91058 Erlangen, Germany}
\author{Eva Weig}
\email{eva.weig@tum.de}
\affiliation{University of Konstanz, Department of Physics, Universitätsstr. 10, 78457 Konstanz, Germany}
\affiliation{Technical University of Munich, Department of Electrical and Computer Engineering, Theresienstr. 90, 80333 München, Germany}

\date{\today}

\begin{abstract}


In recent years, nanomechanics has evolved into a mature field, with wide-ranging impact from sensing applications~\cite{Chaste.2012,Lepinay.2017,Rossi.2017} 
to fundamental physics~\cite{Riedinger.2018,Kalaee.2019,ArrangoizArriola.2019}, 
and it has now reached a stage which enables the fabrication and study of ever more elaborate devices. This has led to the emergence of arrays of coupled nanomechanical resonators as a promising field of research~\cite{Buks.2002,Zalalutdinov.2006,Hatanaka.2014,Huang.2016,Cha.2018}, 
serving as model systems to study collective dynamical phenomena such as synchronization~\cite{Matheny.2014,Zhang.2015} %
or topological transport~\cite{Huang.2016,Cha.2018,Ren.2020,Ma.2021}. From a general point of view, the arrays investigated so far represent scalar fields on a lattice. Moving to a scenario where these could be extended to vector fields would unlock a whole host of conceptually interesting additional phenomena, including the physics of polarization patterns in wave fields and their associated topology. Here we introduce a new platform, a two-dimensional array of coupled nanomechanical pillar resonators, whose orthogonal vibration directions encode a mechanical polarization degree of freedom. We demonstrate direct optical imaging of the collective dynamics, enabling us to analyze the emerging polarization patterns and follow their evolution with drive frequency.


\end{abstract}

\maketitle




When the vectorial character of electromagnetic waves was established in the 19th century, this opened the door to the interpretation of a wealth of important phenomena, launching the field of polarization physics. Surprisingly, the topological nature of spatially inhomogeneous polarization patterns in wave fields was analyzed much more recently~\cite{Nye.1983,Dennis.2009}
, opening a novel domain of inquiry that continues to draw fresh attention and enables modern applications, e.g. in nano-optics~\cite{Flossmann.2008,Bauer.2015,Sala.2015,Angelis.2019}.

In the world of nanomechanical resonators, it is more challenging to observe polarization physics, even at the level of a single resonator. In the mechanical domain, "polarization" refers to the excitation of motion along different directions. Observing nontrivial effects requires that these vibrational modes are at least almost degenerate, i.e. a geometry with a high degree of symmetry is required. Pioneering experimental works have observed two degenerate orthogonal modes with strong coupling within string resonators or nanowires~\cite{Faust.2012b,Lepinay.2017,Rossi.2017}, 
and two nonlinearly coupled modes in a nanowire~\cite{Perisanu.2010}. Going from one or a few such resonators with polarization degrees of freedom to an entire coupled array would enable accessing the wealth of phenomena in polarization fields that have so far only been studied for electromagnetic waves. In recent years, coupled nanomechanical arrays have emerged as a promising platform for observing collective phenomena and transport ~\cite{Matheny.2014,Zhang.2015,Huang.2016,Cha.2018,Ren.2020,Ma.2021}. However, what has been missing so far is a successful integration of polarization degrees of freedom into an array of coupled resonators. 

In view of the goal to observe and study polarization patterns, an important aim (besides large-scale integration and coupling) is the ability to easily visualize the motion, in a spatially resolved way. This rules out stiff resonators such as nanobeams or -strings, which, as a result of their small vibrational amplitudes need to be measured individually by very sensitive optical or electrical means and where imaging could at best be achieved in a slow sequential fashion in a scanning tip approach. 

On the other hand, nanopillar resonators~\cite{Paulitschke.2013,Rossi.2017,Lepinay.2017,Doster.2019,Molina.2020} offer large flexural motion in two orthogonal directions, and 
have thus been proposed~\cite{Fosel.2017} as 
a natural candidate for rapid spatially resolved optical whole-array imaging of polarization patterns.

\begin{figure}[ht]
    \centering
    \includegraphics[width=\columnwidth]{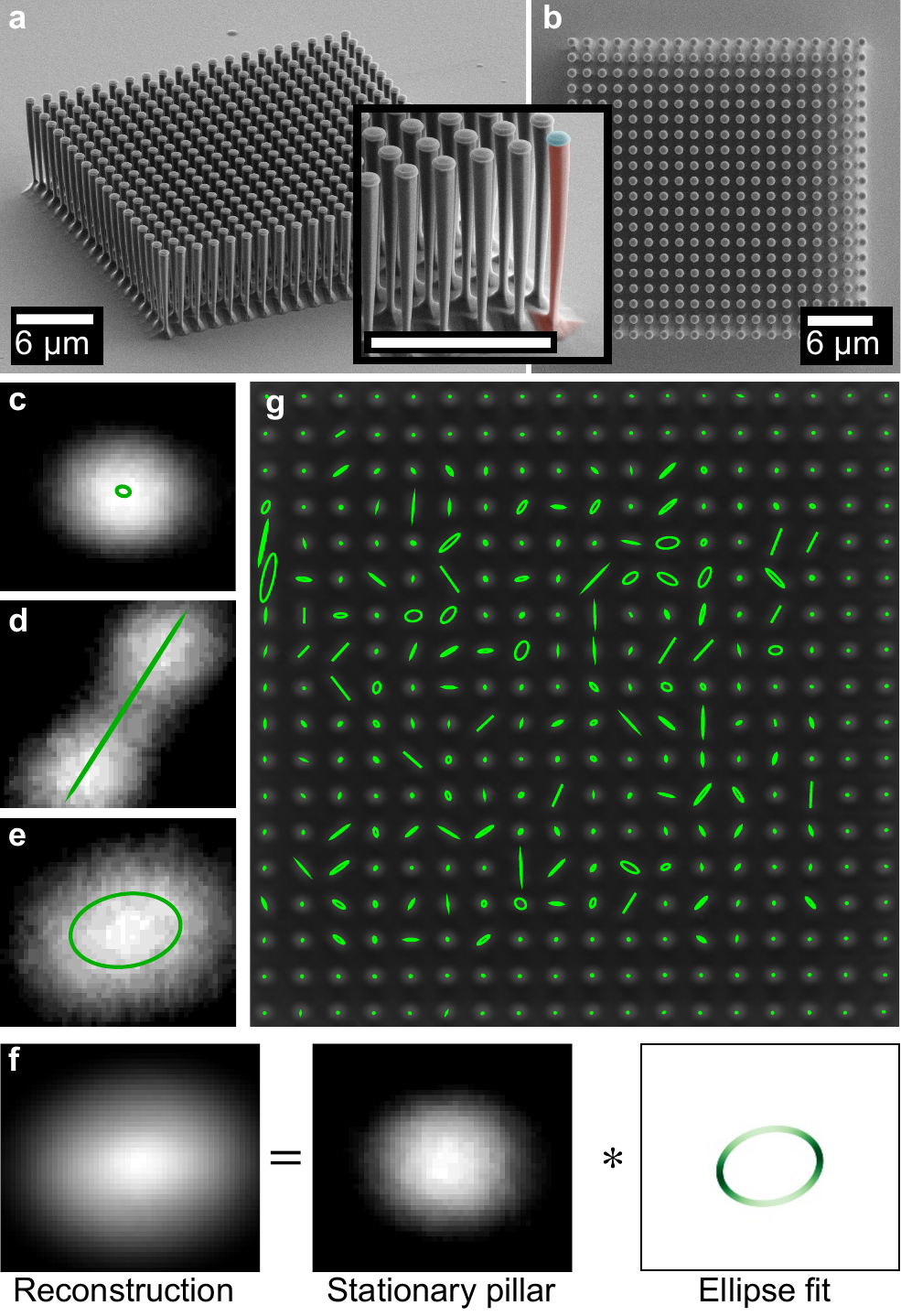}
    \caption{
        \textbf{Characterization of the sample and spatially resolved imaging of vibrational motion.}
        \textbf{a} and \textbf{b}, Scanning electron micrographs of a $20\times20$ nanopillar array with lattice constant \SI{1.4}{\micro\meter}, pillar diameter $d\approx\SI{300}{\nano\meter}$ (measured at the bottom), height $H\approx\SI{6.5}{\micro\meter}$ and taper angle $\varphi\approx\SI{1.5}{\degree}$ in a \SI{60}{\degree} tilted view and in top view. The inset in \textbf{a} shows a zoom of the array corner in the tilted view. False-colors on a single pillar indicate the inverted conical GaAs pillar (red) and the \ce{SiO2} etch mask (blue).
        \textbf{c, d} and \textbf{e} show top view images of three different pillars (extracted trajectories added in green) differentiating between \textbf{c} a pillar at rest and vibrating pillars with \textbf{d} linear trajectory and \textbf{e} elliptical trajectory.
        \textbf{f}, Visualization of the convolution equation. The reconstructed moving pillar image (here, shown for the pillar in \textbf{e}) is a convolution (denoted by *) of the stationary (non-driven) pillar image and the time-average of the fitted elliptical trajectory; details can be found in Appendix~\ref{app:image analysis}.
        \textbf{g}, Top view of the $18\times18$ central pillars of the array in \textbf{a} driven at $\Omega/2\pi=\SI{1.3694}{\mega\hertz}$. The outermost row 
        is omitted here.
    }
    \label{fig:fig12}
\end{figure}

In this work, we investigate an array of $400$ nanomechanical pillar resonators (\autoref{fig:fig12}\textbf{a} \& \textbf{b}). Each nanopillar exhibits two orthogonal fundamental flexural vibration modes with frequencies typically located in the lower \si{\mega\hertz} regime. Coupling between adjacent nanopillars via the strain mediated by the substrate has recently been demonstrated~\cite{Doster.2019}, joining a small number of platforms in which strong coupling of nanomechanical resonators was successfully explored~\cite{Karabalin.2009,Okamoto.2013,Huang.2016,Gajo.2017,Pernpeintner.2018,Cha.2018,Ren.2020,Mathew.2020}. 
The coupling strength can be engineered by adjusting the pillar geometry as well as the separation of the pillars. Here, the geometrical parameters of the array are optimized for both large coupling rate and vibration amplitude.

The nanopillars are driven at a variable frequency $\Omega/2\pi$. Due to the large vibration amplitudes of the pillar heads even in the linear response regime, the envelope of their trajectories can be captured by optical imaging from the top (for details see Appendix~\ref{app:Imaging Setup}). The optical imaging allows for the simultaneous detection of up to several thousands of nanopillars and their spatial trajectories as a function of frequency, whereas typical measurement techniques for resonator arrays rely on sequential measurements of every single resonator~\cite{Molina.2020} or compromise by giving up spatial resolution~\cite{Buks.2002}. 

In the resulting micrographs, a pillar at rest appears as a bright circle (\autoref{fig:fig12}\textbf{c}), whereas a vibrating pillar is swept along its trajectory during the imaging process, yielding the envelope of its motion pattern (\autoref{fig:fig12}\textbf{d-e}). We reconstruct the trajectory by demanding that its convolution with the image of a resting pillar reproduces the observation (\autoref{fig:fig12}\textbf{f}; cf. Appendix~\ref{app:image analysis} for details on the algorithm). The extracted trajectories range from linear to elliptical (\autoref{fig:fig12}\textbf{d-e}).

The variety of motional patterns observed in the whole array (\autoref{fig:fig12}\textbf{g}) indicates a certain amount of disorder. Even as a result of minute geometrical variations arising during fabrication, nanoresonators, though nominally identical, typically show a spread in their eigenfrequencies (Appendices~\ref{app:disorder experiment} and \ref{sec:tight-binding_parameters}). Nonetheless, and despite their narrow linewidth of roughly $\Gamma/2\pi \approx\SI{5}{\kilo\hertz}$ at ambient conditions, a large group of nanopillars vibrates at the same drive frequency. 
This already suggests that the array shows collective motion, which will be demonstrated in more detail later on.




We note that elliptical trajectories are observed despite a linear drive, and we will now briefly describe the physics behind that for a single pillar, before moving on to the dynamics of the entire array. When applying an external drive, it will generally excite both linear polarizations with displacements $x$ and $y$, respectively. In general, due to fabricational anisotropies, these will have different resonance frequencies $\omega_{x,y}$. As discussed in the following, this leads to a phase lag in the response, which can create elliptical motion. 

\begin{figure*}
\centering
\includegraphics[width=2\columnwidth]{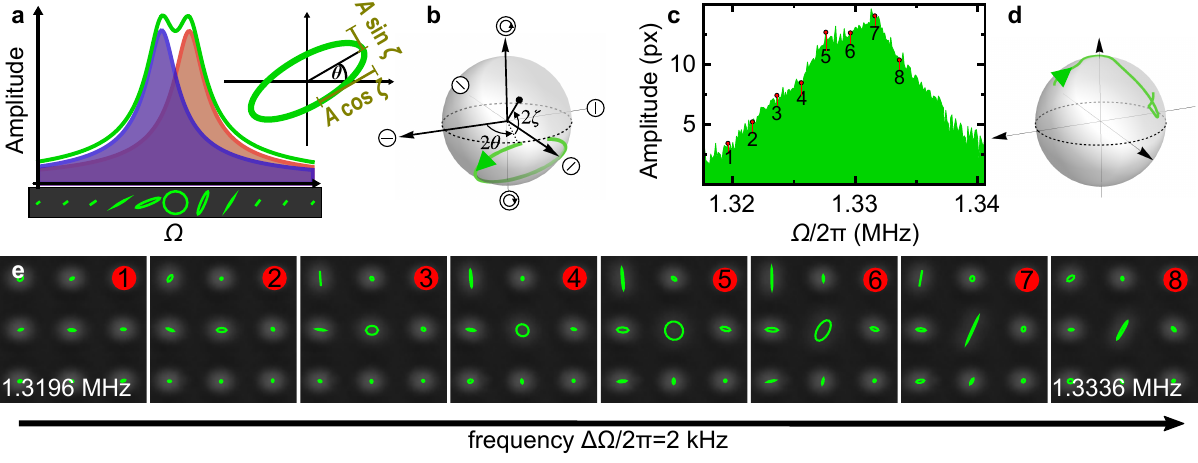}
\caption{\textbf{Polarization physics in a single nanopillar.} 
\textbf{a}, Example of a theoretical frequency response for two separate orthogonal modes of a pillar (blue and red) and the combined response (green). The evolution of the trajectory along the frequency axis is indicated below the diagram. An additional inset shows the nomenclature for an arbitrary elliptical trajectory with semi-major (minor) axis length $A \cos \zeta$ ($A \sin \zeta$), and orientation of the major axis $\theta$.
\textbf{b}, Theoretical trajectory for the example in \textbf{a} on the Poincar\'e sphere. Note that the motion is (counter-)clockwise in the (upper) lower hemisphere.
\textbf{c}, Measured frequency response of a pillar with nearly degenerate modes. The amplitude is expressed in units of a camera pixel, where $1$\,px corresponds to approx. $28$\,nm.
The corresponding path on the Poincar\'e sphere is shown in \textbf{d} (data have been smoothed, see Appendix~\ref{app:image analysis}). Note that we cannot measure the circulation sense of the ellipse in the experiment, therefore the path is depicted on the upper hemisphere for convenience. 
\textbf{e}, Experimental evolution of the pillar's trajectory with drive frequency in the center of the images. The central pillar indicates the transition between the two orthogonal vibration directions via an elliptical trajectory. Frequency steps between neighboring images are $\Delta \Omega /2\pi=\SI{2}{\mega\hertz}$.}
\label{fig:fig3}
\end{figure*}

It is convenient to employ complex notation, 
\begin{equation}
\label{Eq:complex_amplitude}
b_x = \sqrt{\omega_x/2}(x + i {\dot x}/{\omega_x}),
\end{equation}
and likewise for $y$. Then, we will have $b_x = f_x e^{-i \Omega t} / ( (\omega_x - \Omega) - i \Gamma / 2 )$ where $f_x\sim \cos(\varphi)$ is proportional to the force amplitude along $x$, for a linear drive along direction $\varphi$, and likewise for $b_y$ (Appendices \ref{app:single pillar} and \ref{SI:shape_of_ellipse}). Crucially, as we sweep the drive frequency $\Omega$, the phase lag between both linear polarizations (i.e. the phase of $b_y/b_x$) shifts. This leads to a transition from linear polarization to elliptical back to linear, even for a single pillar, as shown in \autoref{fig:fig3}\textbf{a,b}. 

At first sight, it might seem surprising that elliptical motion patterns can emerge in this system, as they are not time-reversal invariant (selecting a sense of circulation), while both the bare model of an anisotropic oscillator and the linear drive itself conserve time-reversal symmetry. This is resolved by noting that the phase lag leading to such motion only arises in the presence of dissipation, which does break time-reversal symmetry.

This theoretical description is borne out when observing a single pillar within the array (\autoref{fig:fig3}\textbf{c-e}). Both the spectrum  (\autoref{fig:fig3}\textbf{c}) and the Poincar\'e sphere trajectory (\autoref{fig:fig3}\textbf{d}) show deviations from the idealized response of a single pillar, but this can be explained by the influence of the collective modes of the array. 
Apart from this, the overall features of the frequency evolution of the mechanical polarization (\autoref{fig:fig3}\textbf{d,e}) are consistent with two spectrally overlapping linear eigenmodes. 


\label{sec:tight_binding}

\begin{figure*}
    \centering
    \includegraphics[width=2\columnwidth]{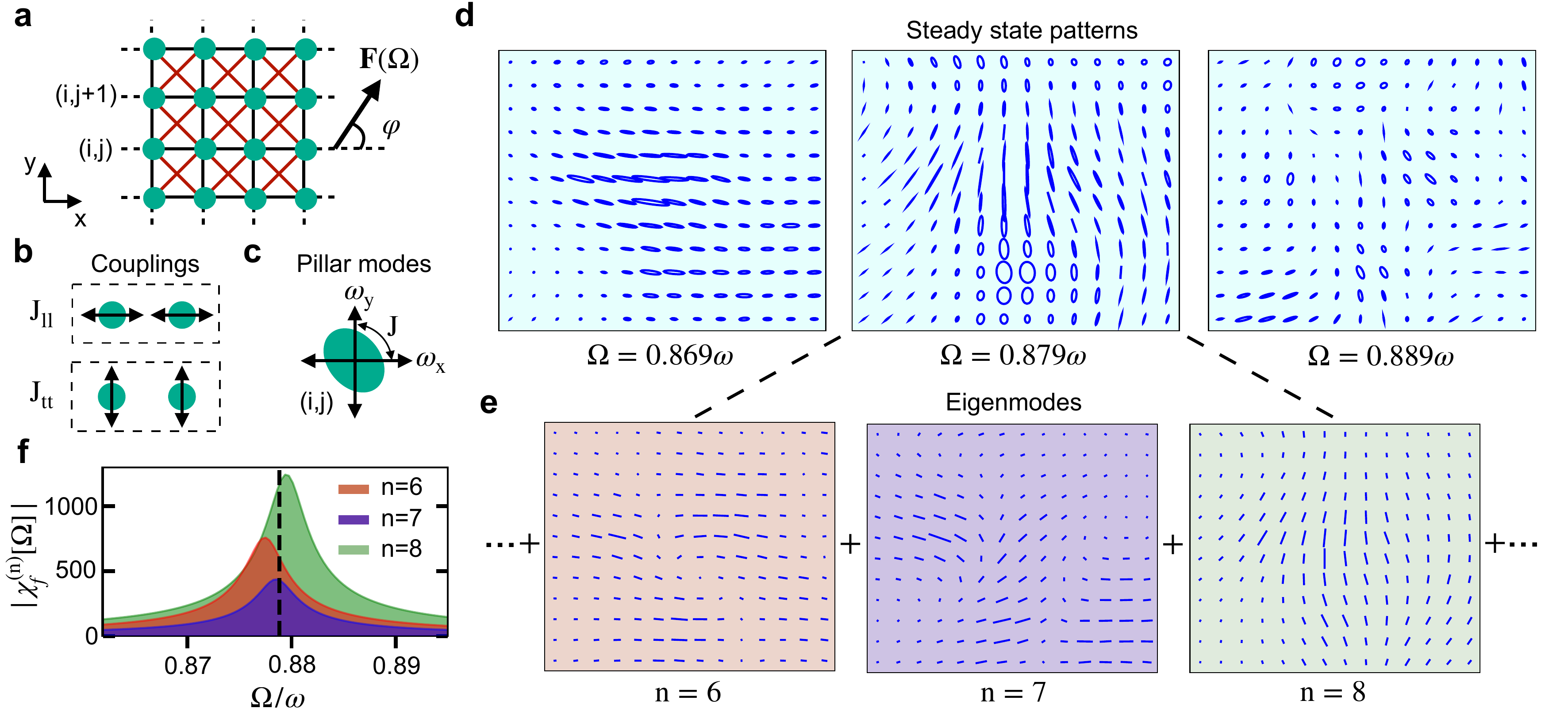}
    \caption{
        \textbf{Tight-binding model for coupled pillar dynamics.}
        \textbf{a}, Schematics of a pillar array with the nearest (black) and the next-to-nearest (red) neighbor couplings. 
        \textbf{b}, Sketch of the two types of couplings between the neighboring pillars. 
        \textbf{c}, Modelling an isolated pillar as two coupled harmonic oscillators with frequencies $\omega_{x,y}$ and coupling $J$. 
        \textbf{d}, Steady state motion in a section of the array for various driving frequencies $\Omega$ renormalised with mean pillar frequency $\omega$. Each pattern is composed of a linear superposition of all the eigenmodes with their appropriate susceptibility $\vert \chi_{f}^{(n)}[\Omega] \vert $. 
        \textbf{e}, Three highest contributing eigenmodes for the steady state pattern at $\Omega=0.879\omega$. Notice that the eigenmodes contain only linear motion, thus the steady state pillar trajectory can be elliptical only if more than one eigenmode contributes to it. 
        \textbf{f}, Contribution of each eigenmode in \textbf{e} to the steady state pattern as a function of $\Omega$. The dashed line indicates $\Omega=0.879\omega$. Parameter values of the tight-binding model are given in Appendix~\ref{sec:tight-binding_parameters}.
    }
    \label{fig:fig4}
\end{figure*}

Based on our analysis of a single pillar and its polarization physics, we can now study the full array. Our theoretical analysis relies on a tight-binding model. In Ref.~\cite{Doster.2019}, it has been shown experimentally that the coupling strength between pillars decreases with distance. Thus, in the model, we only consider the couplings between the nearest (side) and the next-to-nearest (diagonal) neighbors (\autoref{fig:fig4}\textbf{a}).

The interaction between neighboring pillars depends both on the relative vibration direction of the two pillars (see \autoref{fig:fig4}\textbf{b}) 
and their distance. If the two pillars move perpendicular (parallel) to their line of connection, we call the interaction transversal (longitudinal), with coupling strength $J_{\rm tt}$ ($J_{\rm ll}$). 
Arbitrary anisotropies of any pillar can be fully characterized by introducing the frequencies $\omega_{x,y}$ and a coupling $J$ between $x$ and $y$ (cf. \autoref{fig:fig4}\textbf{c}). 


In summary, the Hamiltonian of a $N \times N$ pillar array can be expressed in terms of the complex amplitudes $b_{x,y}$ (see \autoref{Eq:complex_amplitude}) of the individual pillars as 
\begin{align}
\label{eq:ham}
H=& \underbrace{\sum _{s,\mathbf{r}} \omega_{\mathbf{r},s}b^{*}_{\mathbf{r},s}b_{\mathbf{r},s}-J_{\mathbf{r}}b^{*}_{\mathbf{r},s}b_{\mathbf{r},\bar{s}}}_{\text{on-site Hamiltonian}}
-
\underbrace{J_{\rm ll} \sum _{s,\langle \mathbf{r},\mathbf{r}' \rangle_{s}} b^{*}_{\mathbf{r},s}b_{\mathbf{r}',s}}_{\text{n.n longitudinal coupling}} \nonumber \\
&-
\underbrace{J_{\rm tt} \sum _{s,\langle \mathbf{r},\mathbf{r}' \rangle_{\bar{s}}} b^{*}_{\mathbf{r},s}b_{\mathbf{r}',s}}_{\text{n.n transversal coupling}}
+
\underbrace{H_d}_{\text{n.n.n coupling}}.
\end{align}
Here, $\mathbf{r}=(i,j)$ indicates the position of a pillar in the array, $s=\{x,y\}$ labels the direction of motion, and $\langle \mathbf{r},\mathbf{r}' \rangle _s$ indicates the nearest neighbors in the $s$ direction. The bar symbol in the on-site Hamiltonian and the transversal coupling interchanges the two directions i.e. $\bar{x}=y$ and vice-versa. 
For a realistic analysis of the experiment, this model is supplemented by a description of the disorder, as shown in the Appendix~\ref{sec:disorder_theo} (together with the explicit form of the next-to-nearest neighbor coupling terms in Appendix~\ref{app:nnn coupling}).

The steady state response of the array can then be understood by decomposing into contributions from all the eigenmodes, cf. \autoref{fig:fig4}\textbf{d-f} (see Methods).

\begin{figure*}
    \centering
    \includegraphics[width=2\columnwidth]{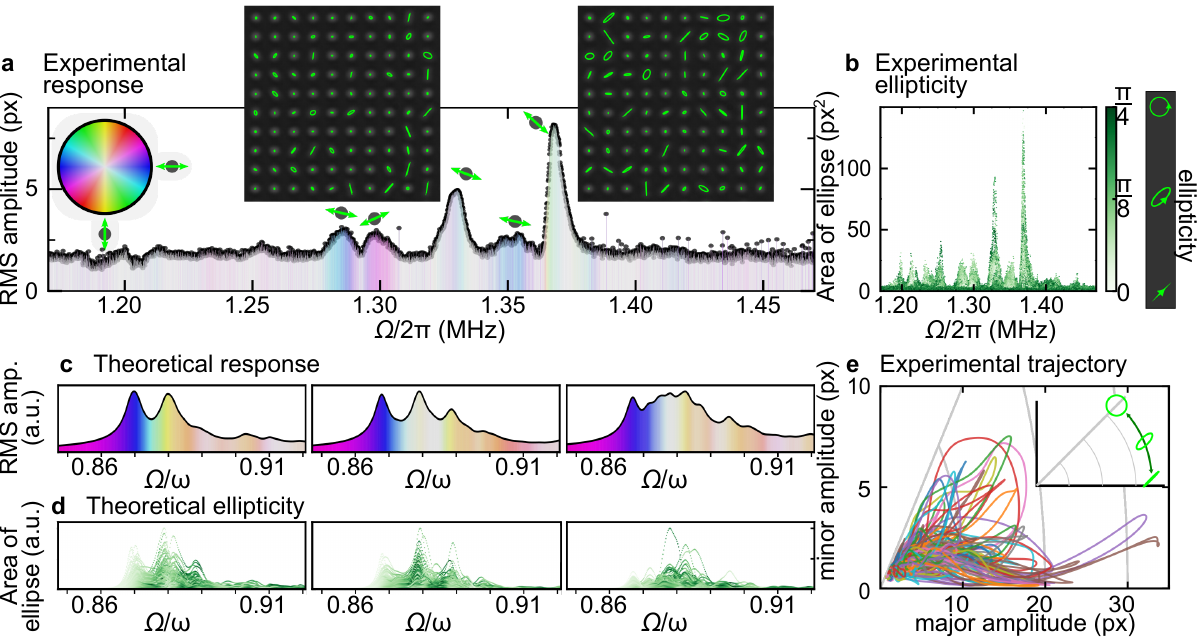}
    \caption{
        \textbf{Evolution of polarization patterns.} 
        \textbf{a}, Measured RMS amplitude for the pillars in the array vs. drive frequency $\Omega/2\pi$. The color wheel illustrates the orientation $\theta$ of the pillar trajectories averaged over the whole array (more intense colors indicate larger homogeneity of the orientation). The small insets at each peak reflect the average orientation at the maximum of the peak. The experimental steady state patterns at the two largest peaks are shown as further insets.
        \textbf{b}, Experimental area of the elliptical trajectory for different drive frequencies with color-coded ellipticity $\zeta$. The pictograms next to the color bar illustrate how ellipticity corresponds to the actual form of the trajectory. 
        \textbf{c}, Theoretical RMS amplitude 
        for different frequencies displayed in three different random realizations with the same disorder parameters. The color reflects the mean orientation of the array at each frequency according to the color wheel in \textbf{a}. 
        \textbf{d}, Theoretical area of the elliptical trajectory depending on the drive frequency with color coded ellipticity as in \textbf{b} for the three disorder realizations in \textbf{c}. 
        \textbf{e}, Experimentally determined minor amplitude vs major amplitude of each pillar in the array around the largest peak: ($\SI{1.36}{\hertz} \leq \Omega/2\pi \leq \SI{1.38}{\hertz}$; data have been smoothed, see Appendix~\ref{app:image analysis}). As shown in the inset, trajectories become more elliptical when moving up from the $x$-axis to the diagonal. 
        Tight-binding parameter values corresponding to \textbf{c} and \textbf{d} are given in Appendix~\ref{sec:tight-binding_parameters}. 
    }
    \label{fig:fig5}
\end{figure*}

With this theoretical model in hand, we can now study the experimentally observed frequency-dependent polarization patterns of the array, where we focus on the central $18\times18$ pillars, to avoid boundary effects (see Appendix~\ref{app:disorder experiment}). This is in contrast to \autoref{fig:fig3}, where we were interested in the dynamics of a single pillar.

In \autoref{fig:fig5}\textbf{a}, we show the experimentally observed steady-state patterns and the RMS amplitude $\sqrt{\sum_{\mathbf{r}}A_{\mathbf{r}}^2}/N$ as a function of the drive frequency $\Omega$. The amplitude response peaks at certain frequencies, as opposed to observing an uninterrupted band extending over all the eigenfrequencies of the array. This is because only eigenmodes with predominantly long-wavelength contributions couple constructively to the uniform drive such that only the lower end of the frequency band and hence its first few modes are experimentally accessible. 
In addition, the two strongest peaks 
feature elliptical pillar motions (see insets of \autoref{fig:fig5}\textbf{a}), hence according to our earlier analysis there must exist at least two (linearly polarized) array eigenmodes within the bandwidth of these peaks. A well-established useful quantity in polarization physics is the complex Stokes field $\sigma=A^2 \cos (2\zeta) e^{2i\theta}$ (see  \autoref{fig:fig3}\textbf{a,b}). By studying its average across all pillars, ${\sum_{\mathbf{r}}\sigma_{\mathbf{r}}} / {\sum_{\mathbf{r}}A_{\mathbf{r}}^2}$,  we can extract both the mean orientation (via the phase) and its fluctuations (via the magnitude). In \autoref{fig:fig5}\textbf{a}, these quantities have been color-coded to illustrate the evolution with frequency.



We now go beyond average quantities and study the distribution of individual ellipticities $\zeta$ across all the observed $18^2$ pillars. The resulting scatterplot (\autoref{fig:fig5}\textbf{b}) reveals that the majority of elliptical trajectories are observed at the two strongest resonances.
It is equally illuminating to track the frequency-evolution of attributes like minor and major axis of each pillar \autoref{fig:fig5}\textbf{e}, which clarifies that strongly elliptical motion is confined to a handful of pillars only.

It is not practicable to extract the (large) number of tight-binding model parameters from the experimental data, but fortunately many of our observations can still be qualitatively captured very well by the theoretical model. The effects of disorder are illustrated very well by running numerical simulations on nominally identical parameters, but for different disorder realizations (\autoref{fig:fig5}\textbf{c,d}). On the one hand, this demonstrates significant sample-to-sample fluctuations, but on the other hand, robust features can be identified. For instance, in agreement with the experimental observations, some of the resonances are primarily linearly polarized, while others support the elliptically polarized motion patterns discussed above. 

All in all, the findings of \autoref{fig:fig5} convincingly demonstrate the existence of collective motional polarization patterns in the nanopillar array. The observed vibrational patterns can not be explained by the independent co-vibration of individual pillars, but require the coupling between adjacent pillars of the array.



\begin{figure}
\centering
\includegraphics[width=1\columnwidth]{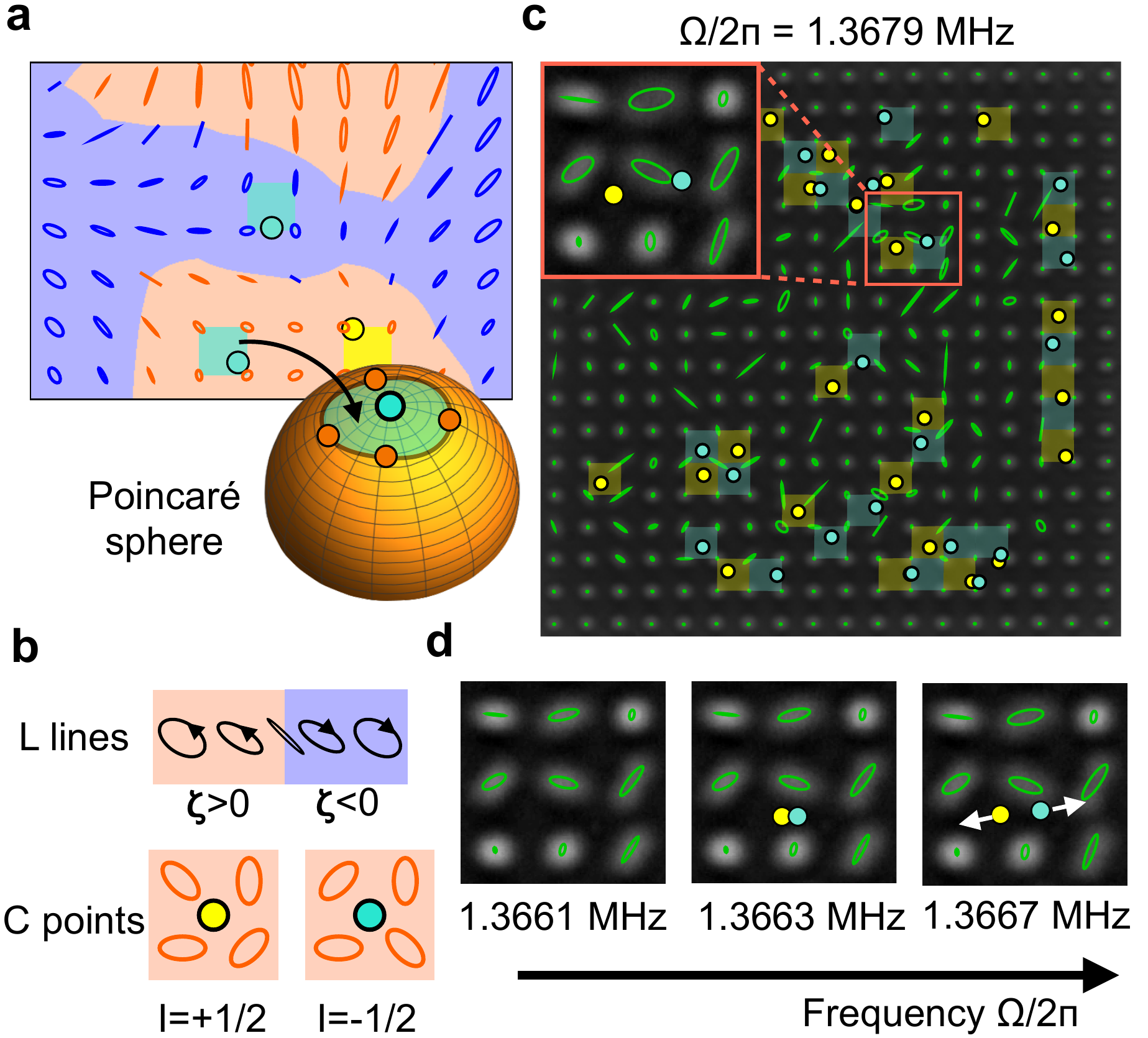}
\caption{\textbf{Topological singularities in mechanical polarization patterns.} 
\textbf{a-b}, Illustration of L lines and C points as topological singularities in a simulated motion pattern (with disorder level reduced compared to the experiment for easier visualization). A plaquette likely contains a C point if it maps to a patch on the Poincar\'e sphere that encloses the pole. \textbf{c}, Experimentally observed pattern with algorithmically suggested candidates of C points (see Appendix~\ref{app:topo singularities}). Their topological winding index $I$ (winding of the orientation $\theta$ around the C point) is indicated by color. \textbf{d}, A frequency sweep reveals a situation where a pair of C points of opposite index are created and move apart. Note that the algorithm achieves sub-lattice-spacing resolution via interpolation, and estimated trajectory data have been smoothed in \textbf{c} and \textbf{d}, (see Appendix~\ref{app:image analysis}).}
\label{fig:fig6}
\end{figure}

By virtue of setting up hundreds of coupled nanopillars in an array and the fast optical imaging measurement technique, our system opens the door to the general area of complex spatial polarization patterns. This includes the investigation of robust topological features~\cite{Nye.1983,Dennis.2009}: 
the ‘L lines’ (curves of linear polarization) and ‘C points’ (isolated points of circular polarization; as seen in \autoref{fig:fig3}\textbf{e}, panel 5) in the vibration patterns of the nanopillars (\autoref{fig:fig6}\textbf{a,b}). We can extract the locations of C points by investigating the winding of the ellipse orientation $\theta$ around any plaquette of the array. This technique is able to suggest the location of C points, as seen in \autoref{fig:fig6}\textbf{c}, even without access to the handedness of the motion. We overcome the challenges posed by the disorder and measurement noise via careful data analysis (Appendices \ref{app:image analysis} and \ref{app:topo singularities}). Moreover, we can observe the evolution of C points while sweeping the drive frequency, discovering a pair-creation event of C points of opposite topological index $I=\pm 1/2$. In the future, stroboscopic imaging could reveal the handedness of the motion, which would allow reliable detection of L lines as well.




In summary, we have observed polarization patterns with signatures of collective dynamics in a two-dimensional array of coupled nanomechanical pillar resonators. Our measurements have been enabled by a whole-array optical imaging approach that allowed us to track the evolution of motional patterns with drive frequency. The platform introduced here enables the exploration of polarization fields in nanomechanics, unlocking phenomena for the domain of mechanics that in the past few years have led to a great number of insights and applications in electromagnetic systems. We have discovered first indications of topological singularities in mechanical polarization fields.

The platform's flexibility in fabrication will naturally allow for the implementation of more complex lattice geometries, such as variants of a honeycomb structure, which could be used to study topological transport (e.g. in the valley Hall effect) and its interplay with collective polarization physics. In this context, individual actuation of pillars could be achieved for example by a photothermal drive. 
This would then allow to observe the propagation of wave packets through the array or along edge channels. As with any array platform there can of course be considerable disorder effects, but we managed to reduce these by careful control of the fabrication. Further improvements would enable more detailed studies of topological polarization phenomena. On the other hand, one might choose to focus on issues like Anderson localization, by deliberately increasing the disorder strength.

In future experiments the nanopillars can be even more strongly driven, which would permit the exploration of the collective motion in the nonlinear regime. 
In a different vein, coupling the array to electrical circuits or optical modes promises to both explore alternative sensing techniques as well as optomechanical manipulation, up to and including the excitation of limit cycles composed of collective polarization patterns.









\appendix
\section*{Methods}

\subsection*{Fabrication Details}
\label{app:Fabrication Details}
The conically inverted GaAs nanopillars (cf. \autoref{fig:fig12}\textbf{a}) are fabricated in a top-down fabrication process from a (100) GaAs wafer. The two-dimensional pattern of the array is defined via electron-beam lithography. This allows for a dense spatial integration of the pillars and a high control over the array geometry. A subsequent chlorine-based anisotropic reactive-ion etch with protective etch mask of \ce{SiO2} yields an array of high aspect ratio nanopillars.

\subsection*{Imaging Setup}
\label{methods:Imaging Setup}
We measure the pillars' response to an external drive at room temperature and atmospheric pressure. The external periodic force is applied by a shear piezo glued underneath the sample. The response of the pillars to this drive is then imaged from above the sample (see Appendix~\ref{app:Imaging Setup} for more details). Resting pillars are identified as bright circles. Moving pillars appear smeared out compared to the resting pillar. The image of a moving pillar captures the envelope of its vibrational motion, as the exposure time of the camera, which is in the range of a second, greatly exceeds the oscillation period.

\subsection*{Dynamical Response}
\label{Methods:dynamical_response}
All pillars in the array are subjected to an identical harmonic drive at frequency $\Omega$ and angle $\varphi$ with respect to the $x$-axis. Thus, the driving rate $f_x$ $(f_y)$ is proportional to $\cos \varphi$ $(\sin \varphi )$. For a mechanical damping $\Gamma$ (assumed identical for all pillars), the equation of motion of a pillar in the array is given by
\begin{equation}
\label{eom_array}
\frac{db_{\mathbf{r},s}}{dt}=-i \frac{\partial H}{\partial b^*_{\mathbf{r},s}}-\frac{\Gamma}{2}b_{\mathbf{r},s}+if_{s}e^{-i\Omega t}.
\end{equation}
Here, the partial derivative $\partial H /\partial b^*_{\mathbf{r},s} $ is taken only over $b^*_{\mathbf{r},s}$, while $b_{\mathbf{r},s}$ is held as a constant. The steady state solution can be written as a superposition of all the eigenmodes $b^{\rm (n)}_{\mathbf{r},s}$ (eigenmode index labelled by n) of the Hamiltonian $H$ (cf. \autoref{fig:fig4}\textbf{d},\textbf{e}) as
\begin{equation}
\label{steady_state}
b_{\mathbf{r},s}=\sum _{n=1} ^{2N^2} \chi^{\rm (n)}_f\left[\Omega\right] b^{\rm (n)}_{\mathbf{r},s}  e^{-i\Omega t}, 
\text{  } 
\chi^{\rm (n)}_f[\Omega]=\frac{\sum_{s',\mathbf{r}'} b^{\rm (n)*}_{\mathbf{r'},s'} f_{s'}}{(E_n - \Omega) - i\Gamma /2}.
\end{equation}
Here, the mechanical susceptibility $\chi^{\rm (n)}_f[\Omega]$ depends on both the drive frequency $\Omega$ and the overlap of the eigenmode with the drive (cf. \autoref{fig:fig4}\textbf{f}). 

\bibliography{bib_pillar-arrays}

\noindent\textbf{Acknowledgements}\\ 
The authors gratefully acknowledge technical support from H. Lorenz and P. Paulitschke in the reactive ion etching of the nanopillar arrays at LMU Munich. We further thank P. Paulitschke for valuable discussions about the direct imaging of the dynamics of arrays. T.S. acknowledges support from the European Unions Horizon 2020 research and innovation programme under the Marie Sklodowska-Curie grant agreement No. 722923 (OMT). F.M., J.D. and E.M.W. acknowledge support from the European Union’s Horizon 2020 Research and Innovation program under Grant No. 732894, Future and Emerging Technologies (FET)-Proactive Hybrid Optomechanical Technologies (HOT). J.D. and E.M.W. acknowledge funding from the German Federal Ministry of Education and Research through contract no. 13N14777 funded within the European QuantERA cofund project QuaSeRT.

\clearpage
\onecolumngrid
\appendix

\section{Imaging Setup}
\label{app:Imaging Setup}
We employ optical detection by means of microscopy to simultaneously detect the dynamics of every nanopillar within the array. The array is imaged from above, which allows to capture the envelope of each pillar's trajectory, resolving not only its amplitude but also the vibration direction, which gives access to the polarization degree of freedom. This measurement technique is enabled by the relative large vibrational amplitudes of nanomechanical pillar resonators even in the linear response regime.

\begin{figure}[h]
    \centering
    \includegraphics[width=8cm]{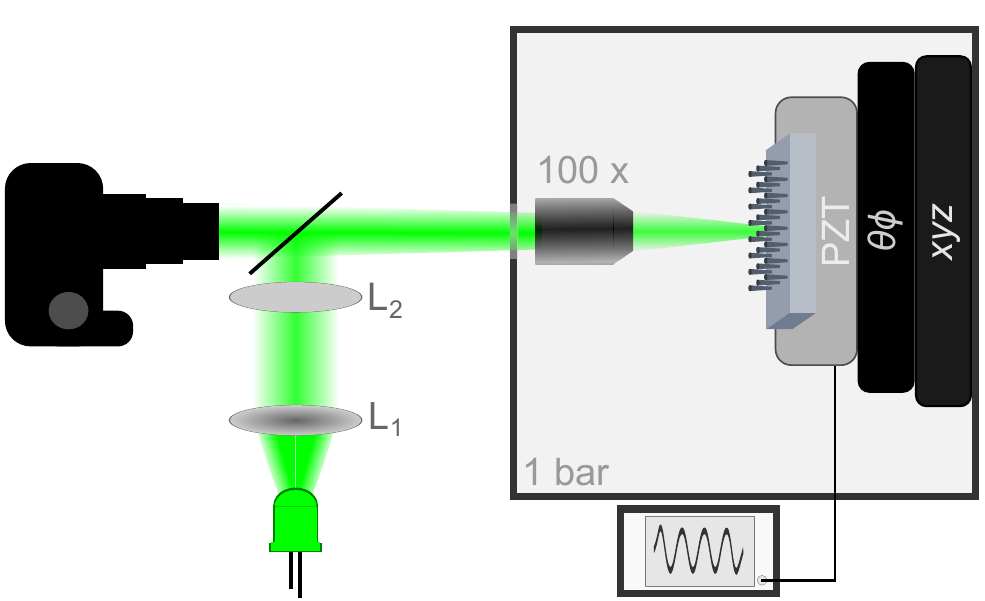}
    \caption{Schematic of the imaging setup for the detection of nanomechanical pillar arrays. Light from a green LED is diffused via a lens L1 exhibiting a grained surface and collimated via a second lens L2. The light is then focused onto the sample via a $100\times$ objective. The light reflected off the sample passes through a beamsplitter and is captured in a DSLR camera. The sample is driven via a piezoelectric transducer (PZT), that is connected to a signal generator. A combination of a $xyz$- and a $\theta\phi$-stage are employed for precise sample positioning.}
    \label{fig:setup}
\end{figure}

A schematic drawing of the experimental setup is shown in Figure~\ref{fig:setup}. The imaging detection system consists of a home-built microscope operating with monochromatic light to avoid imaging distortions from chromatic abberation. The light of a green LED is focused on the pillar array from the top through a $100\times$ magnification objective. A diffuser lens with a grained surface (L1) is employed to homogenize the illuminating light across the large field of view (approx. $100\times100\,\si{\micro\meter\squared}$) required to capture the pillar array. In addition, $2\,$inch optic components are employed to ensure the uniform illumination of the array. The light reflected from the sample is then captured via a digital single-lens reflex (DSLR) camera. Green light is chosen to exploit the spatial sensitivity maximum of the camera chip. 

The sample is positioned with a remote controlled $xyz$-stage and a mechanical $\theta\phi$-tilt-stage. In particular, precise tilt correction is required to tune the entire field of view into focus. This is crucial to discriminate both idle and vibrating pillars, the enveloples of which appear as blurred circles and shapes (see Figs.~1\textbf{c-e} of main text), against unfocussed ones.  
To compensate for position or focus drifts, that might influence the interpretation of the images, reference images with the drive switched off are taken before every image capture of the driven sample.

The nanopillars of the array are driven via a shear piezoelectric transducer glued underneath the sample chip. As the single image capture time of the camera is approx. $1$\,s and thus much larger than the oscillation period of the pillars at eigenfrequencies of about $1$\,MHz, the trajectories are not time-resolved but rather the integrated trajectory is obtained, yielding an image of the resulting envelopes.

All measurements in this article are taken at room temperature and atmospheric pressure.

\section{Extracting motion from the blurred pillar image}
\label{app:image analysis}

\begin{figure}[h]
    \centering
    \includegraphics[width=1\columnwidth]{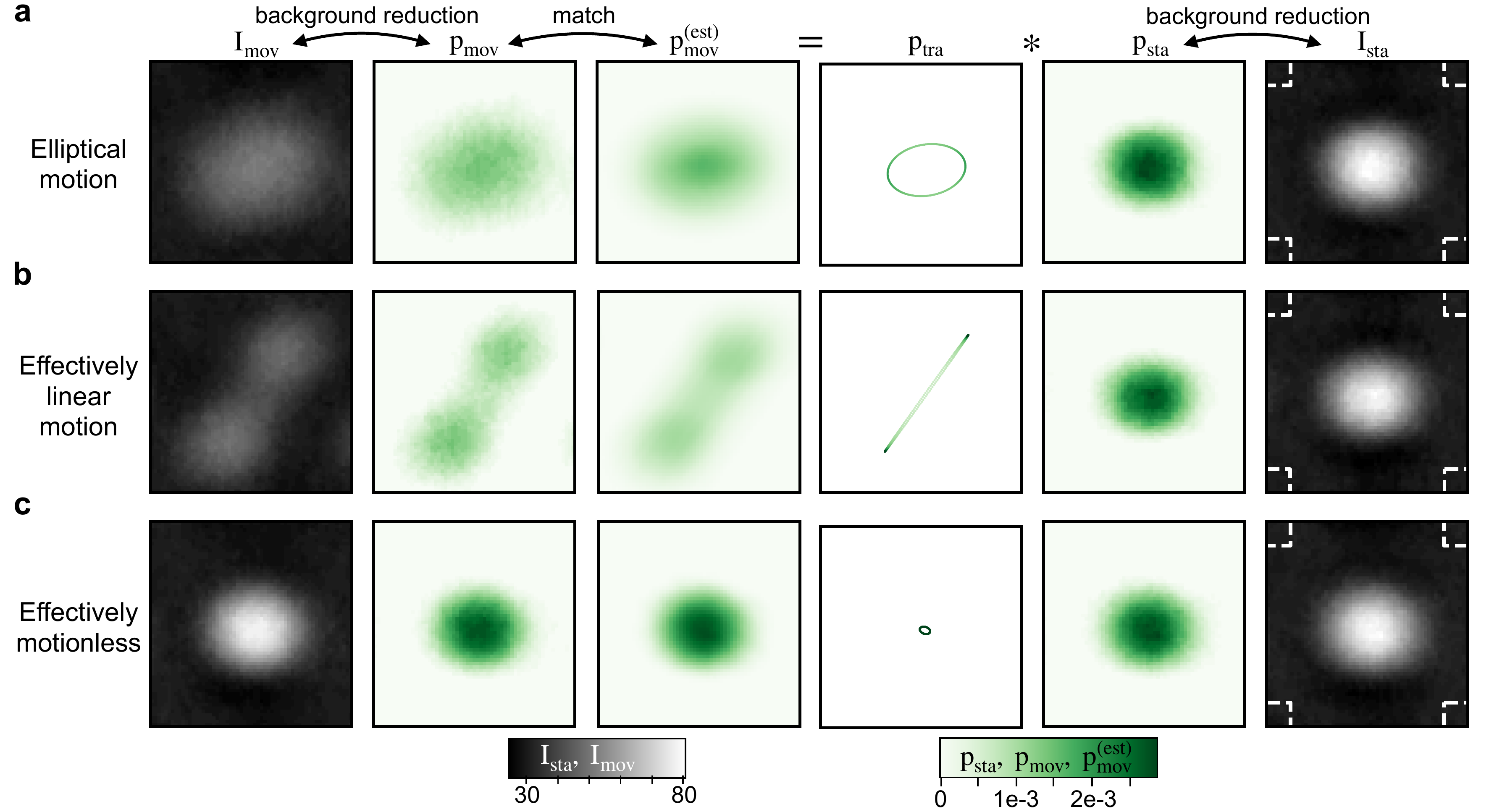}
    \caption{{\bf Image processing algorithm to extract pillar trajectories from the blurred pillar images.} 
    Demonstration of the algorithm with qualitatively different trajectories:
        \textbf{a}, elliptical motion, \textbf{b}, effectively linear motion, and \textbf{c}, effectively motionless.
        The first (last) two columns show the pixel intensity of the moving $I_{\rm mov}$ (stationary $I_{\rm sta}$) pillar and the corresponding background reduced probability distribution $p_{\rm mov}$ ($p_{\rm sta}$); subsequent three columns depict the convolution equation $p_{\rm mov}^{\rm (est)} = p_{\rm sta} \ast p_{\rm tra}$; the estimated distribution $p_{\rm mov}^{\rm (est)}$ is compared with the moving pillar distribution $p_{\rm mov}$ to yield the overlaps 0.992, 0.969, 0.999 for the three cases (a-c). The background pixels $\mathbf{r_{\rm bg}}$, described in Step 1 of the algorithm, are indicated with dashed lines at all the corners of the static pillar image in the last column.
    }
    \label{fig:SI_image_analysis}
\end{figure}

\begin{figure}
    \centering
    \includegraphics[width=1\columnwidth]{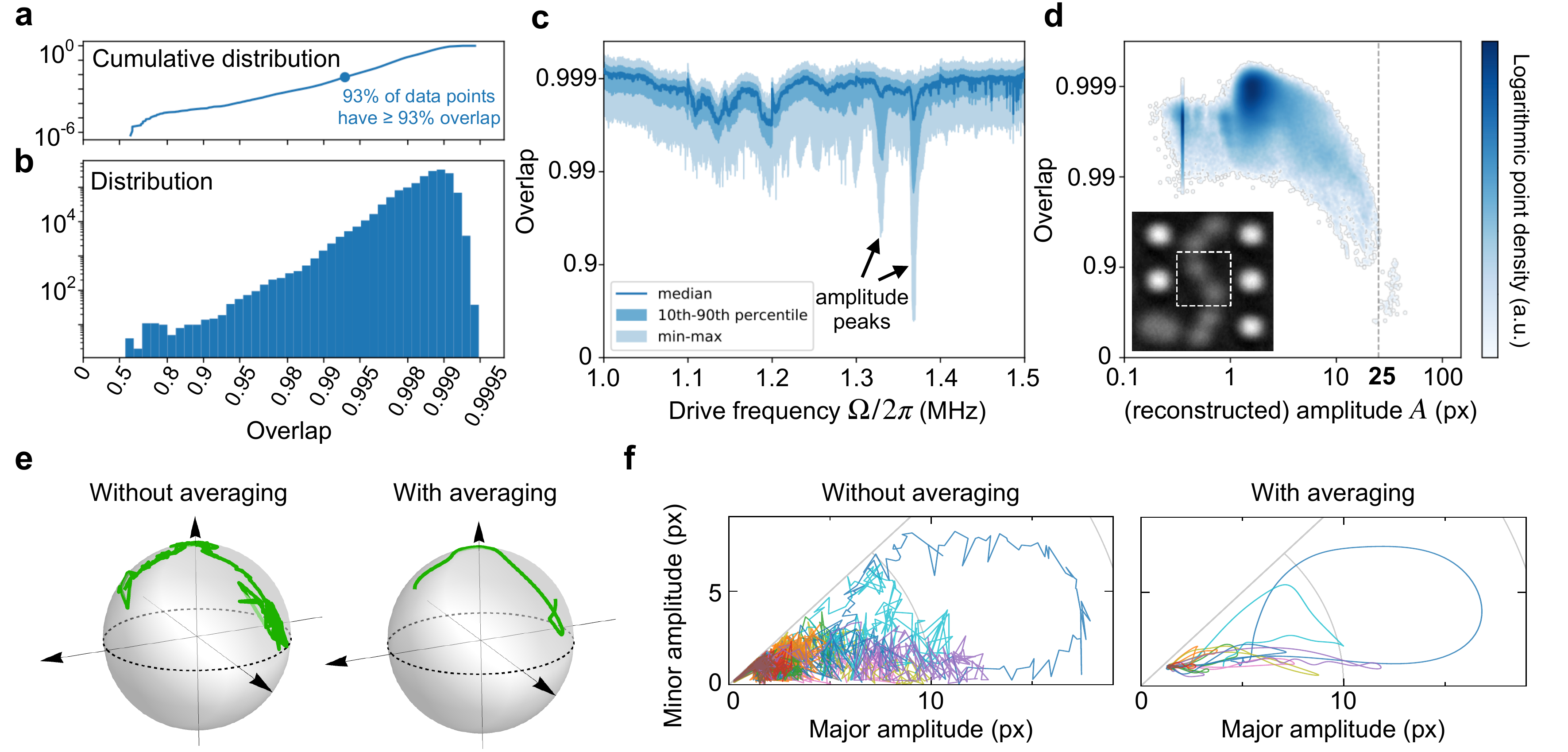}
    \caption{{\bf Important caveats of the estimated trajectories.} 
        \textbf{a-b}, Cumulative distribution and histogram of the overlap $O$ over all the $18 \times 18$ pillars and all driving frequencies $\Omega$. 
        \textbf{c}, Overlap $O$ as a function of the driving frequency $\Omega$. It features dips at the frequencies of maximum amplitude response. The large difference between the median (dark blue) and the minimum value (light blue) at the two largest amplitude peaks indicates that only a minor fraction of the pillars have smaller overlaps. 
        \textbf{d}, Overlap $O$ as a function of the reconstructed amplitude $A$. The overlap decreases with larger amplitude, thereby explaining the dips in \textbf{c}. For amplitudes larger than half the unit-cell size (dashed vertical line), the overlap is worst because either the analysed pillar trajectory or its neighboring pillar trajectories can cross the cropping region (the inset shows such an example with cropping region of the central pillar indicated in white). 
        \textbf{e} (\textbf{f}), Frequency evolution of the estimated single (multiple) pillar trajectory on the Poincar\'e sphere of Figure~2 (major vs. minor amplitude plot of Figure~4) of the main text with and without the Gaussian filter averaging over neighboring frequencies. 
    }
    \label{fig:SI_image_analysis_caveats}
\end{figure}

In this section, we outline the image processing algorithm that is used to extract the pillar trajectory from the stationary and moving pillar images. The underlying idea behind the algorithm is presented in Figure~1\textbf{f} of the main text. It states that the moving pillar image is reconstructed as the convolution of the stationary pillar image with the time-spent probability distribution of the fitted elliptical trajectory. The fitting procedure works by optimizing the ellipse trajectory parameters to match the reconstructed moving pillar image with the experimentally observed one.

Our goal is to determine the ellipse parameters $\boldsymbol{\lambda}=(x_0,y_0,A\cos\zeta,A\sin\zeta,\theta)$ of the ellipse trajectory, where ($x_0,y_0$) is the center position of the ellipse, $\{A\cos \zeta,A \sin\zeta\}$ are the lengths of its two main axes, and $\theta$ is the orientation of the major axis (cf. Fig.~2\textbf{a} of the main text):
        \begin{equation}
            \mathbf{r}(t) =
            \begin{pmatrix}x_0\\y_0\end{pmatrix} +
            A
            R(\theta)
            \begin{pmatrix}\cos\zeta\cos(\Omega t)\\\sin\zeta\sin(\Omega t)\end{pmatrix}.
        \end{equation}
        Here, $R(\theta)$ is the rotation matrix given by 
        \begin{equation}
            R(\theta) = 
            \begin{pmatrix}
                \cos \theta &  -\sin \theta \\
                \sin \theta & \cos \theta
            \end{pmatrix}.
            \label{rot_matrix}
        \end{equation}
For this purpose, we apply the following steps:
\begin{enumerate}
    \item First, we encode the static (moving) pillar shape in a probability distribution $p_{\rm sta(mov)}(\mathbf{r})$. This distribution is defined within a fixed cropping region (here, 1.2 $\times$ unit-cell size) centered at the rest position of the pillar. To ensure that the distribution decays to zero away from the pillar, it is evaluated by subtracting and clipping the gray-scale pixel intensity $I_{\rm sta(mov)}(\mathbf{r})$ by the maximum pixel intensity in the background (away from the pillars, cf. Figure~\ref{fig:SI_image_analysis}) region,
        \begin{equation}
            p_{\rm sta(mov)}(\mathbf{r}) = \mathcal{N}_{\rm sta(mov)} \left[ I_{\rm sta(mov)}(\mathbf{r})-\max(I_{\rm sta}(\mathbf{r_{bg}})) \right]^+.
            \label{calc_p(x)}
        \end{equation}
        Here, $I_{\rm sta}(\mathbf{r_{bg}})$ is the gray-scale pixel intensity of the static pillar at the four corners of the unit cell (see Figure~\ref{fig:SI_image_analysis}), and $\mathcal{N}_{\rm sta(mov)}$ is the normalisation constant. The positive part of a function $[f(\mathbf{r})]^+ \equiv {\rm max} \{ f(\mathbf{r}), 0 \}$ ensures that the probability distribution is always positive. An alternative approach could be to relax the constraint $p_{\rm sta(mov)} \geq 0$, and instead take the average of the background pixel intensities (in alternative to maximum with clipping) for subtraction.
    \item 
        The time-spent probability distribution of the elliptical trajectory is written as 
        \begin{equation}
            p_{\rm tra}(\mathbf{r}) = \frac{\Omega}{2\pi} \int_{0}^{2\pi/\Omega} \delta(\mathbf{r}-\mathbf{r}(t))\,\mathrm{d}t
        \end{equation}
        The center coordinates $x_0$ and $y_0$ are initialized via the mean position, whereas the other parameters are initialized via the covariance matrix, cf. Appendix~\ref{SI:shape_of_ellipse} (mean and variances are evaluated using the moving pillar probability distribution $p_{\rm mov}(\mathbf{r})$).
    \item Next, we estimate the distribution of the moving pillar image $p_{\rm mov}^{\rm (est)}(\textbf{r};\boldsymbol{\lambda})$ via the convolution relation
        \begin{equation}
            p_{\rm mov}^{\rm (est)}[\boldsymbol{\lambda}](\textbf{r}) = (p_{\rm sta}\ast p_{\rm t})(\mathbf{r}) = \int d\mathbf{r'} \text{ } p_{\rm sta}(\mathbf{r} - \mathbf{r'})  \text{ }p_{\rm tra}[\boldsymbol{\lambda}](\mathbf{r'}).
        \end{equation}
        The ellipse parameters $\boldsymbol{\lambda}$ are iteratively optimized to minimize the mean squared error
        \begin{equation}
            \int_{\mathbb{R}^2} [p_{\rm mov}^{\rm (est)}(\mathbf{r})-p_{\rm mov}(\mathbf{r})]^2\,\mathrm{d}^2r
            \label{Eq:overlap_function}
        \end{equation}
        We use gradient descent (Hessian method) for optimization, whose advantage is that it converges very rapidly within at most 5 iterations.
\end{enumerate}

The algorithm works as expected for the three qualitatively diverse test-cases in Figure~\ref{fig:SI_image_analysis}. However, one should keep note of the following important caveats of the estimated trajectories:
\\ [-0.3cm] \par 
(i) Image analysis performance for different amplitudes: We characterize the performance of the image analysis by investigating the optimised overlap function $O=\int p_{\rm mov}^{\rm (est)}(\mathbf{r}) p_{\rm mov}(\mathbf{r}) d^2r$ in Figure~\ref{fig:SI_image_analysis_caveats}\textbf{a-d}. The distribution of the overlap function reveals that the reconstructed image is practically identical to the observed image for almost all the cases, cf. Figure~\ref{fig:SI_image_analysis_caveats}\textbf{a,b}. Additionally, we observe that the large amplitude pillars typically have smaller overlaps (see Figure~\ref{fig:SI_image_analysis_caveats}\textbf{c,d}). This is because at higher amplitudes, the pillar trajectories are very near to (or even outside) the boundary of the cropping region. 
\\ [-0.3cm] \par 
(ii) Improving the data reliability via averaging: A more direct way to test the reliability of the extracted pillar motion is to investigate the frequency evolution of a single pillar trajectory. As can be seen in Figure~\ref{fig:SI_image_analysis_caveats}\textbf{e,f}, the extracted trajectory of the pillar features strong fluctuations, which arises from the combined effects of the data acquisition and the image analysis. These fluctuations can be smoothed-out via a Gaussian averaging of the optimised ellipse parameters over neighboring frequencies.

\section{Characterization of experimental disorder on a different sample}
\label{app:disorder experiment}

\begin{figure}[h]
    \centering
    \includegraphics[width=0.5\columnwidth]{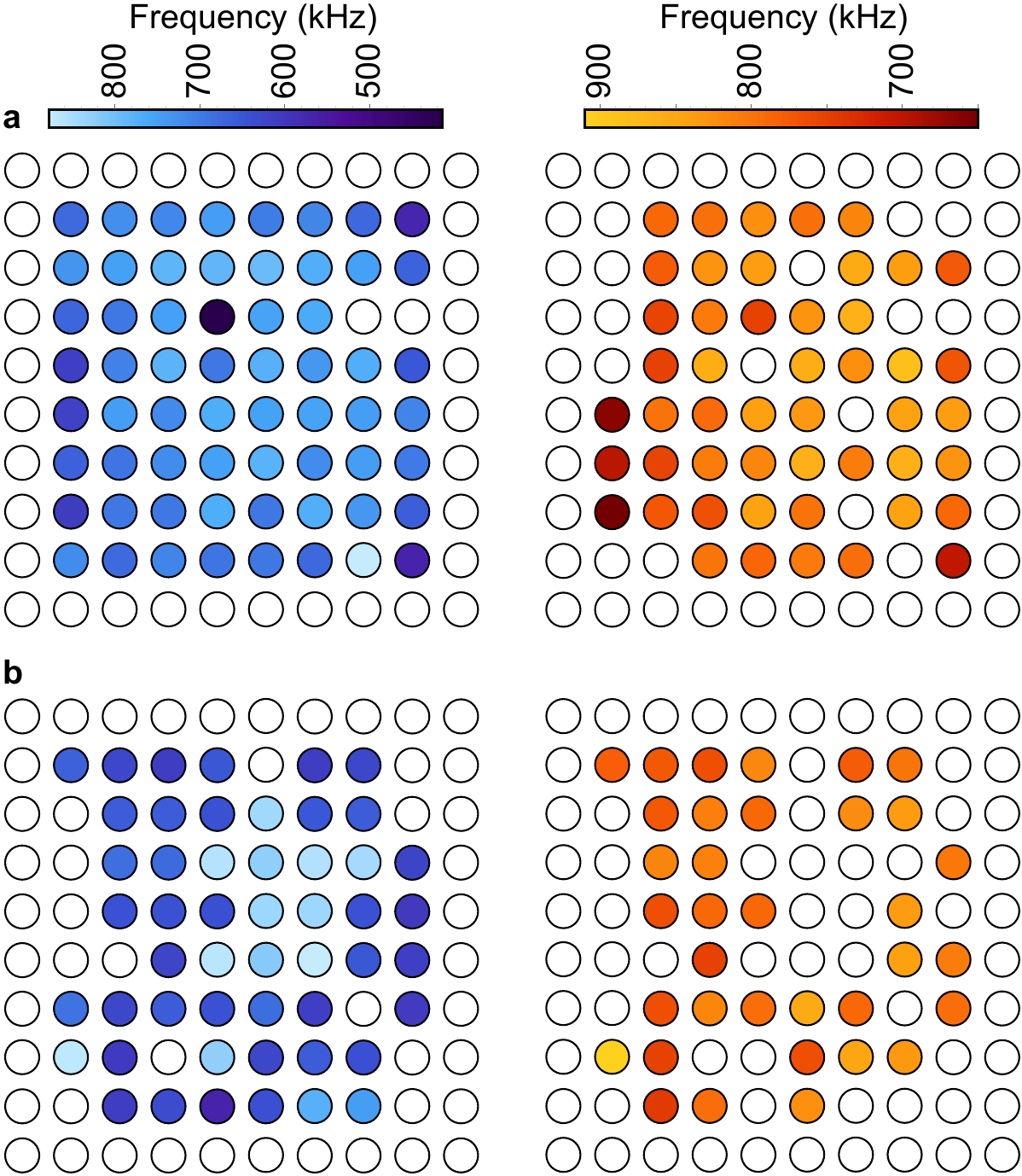}
    \caption{\textbf{Experimental frequency disorder.}
    Frequency distribution for the two main modes of a very weakly coupled array due to a large lattice constant. Two different arrays are displayed in \textbf{a} and \textbf{b}. Both arrays are nominally identical and on the same sample, that is however not the same as the one investigated in the main text but prepared in the same batch. The different modes are shown on the left and the right, respectively, with the same color scale for the two identical arrays. Empty circles indicate pillars whose frequencies lie far from the two main modes.
    }
    \label{fig:disorder_exp}
\end{figure}

The nanopillars in the array are nominally identical. However, despite the fact that the fabrication routines have been optimized to ensure accurate pattern transfer, a certain fabrication-induced disorder in the geometry of nanopillar resonators can not be avoided. In our case, this translates into a disorder in the pillars' resonance frequencies. 
In the nanopillar arrays under investigation, three types of disorder are discerned, which are all apparent in \autoref{fig:disorder_exp}. 

\begin{itemize}
    \item Random disorder between any two pillars in the center of an array: Slight fabricational differences in the radius or height of the pillars can lead to variations of the eigenfrequencies of the two pillars. Optimization of the electron beam lithography and reactive ion etching allows to reduce the underlying artefacts. However, this disorder mechanism can not be fully eliminated and thus needs to be incorporated into the theoretical model. In \autoref{fig:disorder_exp} this can be seen for each mode in the frequency distribution in the center of the array.

\item Length gradient towards the edge of the array: This additional disorder mechanism is based on the diffusion limited reactive ion etching process, for which the etch rate is reduced in highly confined spaces. This essentially leads to a higher etch rate and hence longer pillars towards the edge of the array compared to the center of the array. As the surrounding of all pillars in the central area of the array can be considered effectively the same, systematic deviations are apparent only close to the edge of the array. This is also apparent from \autoref{fig:disorder_exp}. This systematic distortion is the reason, why we neglect the outermost pillar row in all our experimental analyses and effectively only discuss a $18\times18$ rather than the full $20\times20$ array. In consequence, this systematic disorder mechanism is not included into the theoretical model.

\item Anisotropy of the shape of an individual pillar: Although every pillar is nominally written as a circle in the electron-beam lithography process, typically there is some anisotropy to it so that the patterned shape is very slightly elliptical. Again, optimization of the electron beam lithography allows to reduce this effect to a minimal level. However, a certain, barely discernable ellipticity always remains. Typically we find that there is a different preferred direction and strength of this anisotropy randomly changing from sample to sample. This anisotropy appears as a considerable separation between the eigenfrequencies of the two orthogonal vibrational polarization directions in \autoref{fig:disorder_exp}, while the vibration direction within each mode is roughly the same for every pillar (more details in \autoref{sec:tight-binding_parameters}). This disorder mechanism is also included in the theoretical model and introduced in \autoref{sec:disorder_theo}.
\end{itemize}

\section{Theoretical description of the motion of a single pillar}
\label{app:single pillar}
\begin{figure}
\centering
\includegraphics[width=1\columnwidth]{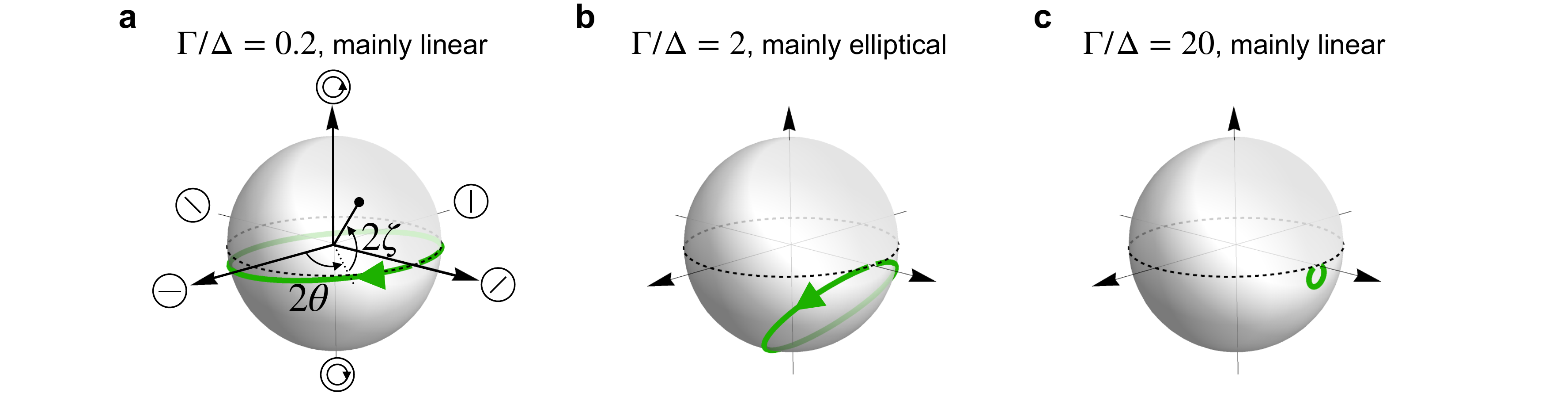}
\caption{{\bf Polarization physics analysis of a single pillar with respect to mechanical damping.}
Evolution of the pillar trajectory on the Poincar\'e sphere as a function of the drive frequency $\Omega$ (frequency increasing in the direction of green arrow) for three different values of the damping $\Gamma$ relative to the frequency anisotropy $\Delta$: \textbf{a}, $\Gamma/\Delta \ll 1$, \textbf{b}, $\Gamma/\Delta \approx 1$, 
\textbf{c}, $\Gamma/\Delta \gg 1$. The trajectory is majorly elliptical (passing close to the poles of the Poincar\'e sphere) only for intermediate values of the damping $\Gamma /\Delta \approx 1$. [Parameter values: $\omega_x=0.95, \omega_y=1.05, \varphi = 45^\circ$.]}
\label{fig:SI_stokes_sphere}
\end{figure}
In Fig.~2 
of the main text, we describe the frequency response of a single pillar and observe its transition from linear polarization to elliptical and back to linear. In this section, we derive the equation of this trajectory, and explicitly illustrate that a non-zero mechanical damping $\Gamma$ is necessary in order to observe elliptical trajectories.

Without any loss of generality, we assume that the two normal modes of the pillar are oriented along the x and y directions, with corresponding frequencies $\omega_x$ and $\omega_y$. In terms of the complex displacement amplitude defined as $b_x = \sqrt{\omega_x/2}(x + i {\dot x}/{\omega_x})$ and likewise for $b_y$, the Hamiltonian is given by
\begin{equation}
\label{single_pillar_hamiltonian}
H = \omega_x b_x^* b_x + \omega_y b_y^* b_y.
\end{equation}
For an external harmonic drive at frequency $\Omega$ and orientation $\varphi$ with respect to the x-axis, the equation of motion is written as
\begin{equation}
\label{eom_single_pillar}
\frac{db_{s}}{dt}=-i \frac{\partial H}{\partial b^*_{s}}-\frac{\Gamma}{2}b_{s}+if_{s}e^{-i\Omega t}, \quad s=\{x,y\}.
\end{equation}
Here, the partial derivative $\partial H /\partial b^*_{s} $ is taken only over $b^*_{s}$, while $b_{s}$ is held as a constant. The force amplitude $f_x$ ($f_y$) is proportional to $\cos \varphi$ ($\sin \varphi$). Thus, the steady state solution is evaluated to be
\begin{equation}
\label{steady_state_single_pillar}
b_s = \frac{f_s e^{-i\Omega t}}{(\omega_s - \Omega) - i\Gamma /2}.
\end{equation}
The shape of the elliptical trajectory can be obtained from this solution according to the procedure described in Appendix~\ref{SI:shape_of_ellipse}.

We now analyse the polarization physics of this trajectory with respect to the mechanical damping $\Gamma$. The phase lag between the two polarizations (phase of $b_y/b_x$) determines the ellipticity. For a drive parked at the central frequency $\Omega = \omega \equiv (\omega_x + \omega_y)/2$, the phase lag is given by
\begin{equation}
\label{phase_lag}
b_y/b_x = \left( \frac{-1 - i\Gamma/(2\Delta)}{1 - i\Gamma/(2\Delta)} \right) \tan \varphi.
\end{equation}
Here, $\Delta = (\omega_y - \omega_x)/2$ is the frequency anisotropy. Note that for the two extreme values of the mechanical damping $\Gamma/(2\Delta) \ll 1$ and $\Gamma/(2\Delta) \gg 1$, the phase of $b_y/b_x$ converges to $\pi$ and $0$ respectively, indicating that the trajectory is mainly linear for these cases. For $\Gamma =0$, the phase of $b_y/b_x$ is exactly $\pi$. Hence, the elliptical trajectories can only be observed by breaking the time-reversal symmetry with finite dissipation. This can be illustrated by visualising the frequency evolution of the pillar trajectory on the Poincar\'e sphere for different values of $\Gamma$, cf. Figure~\ref{fig:SI_stokes_sphere}.
\section{Determining shape of the ellipse from steady state solution}
\label{SI:shape_of_ellipse}
In the main text, we express the steady state solution of a single pillar in the array in terms of the complex amplitudes $b_x$ and $b_y$. In this section, we determine the shape of the elliptical trajectory from this steady state solution.

First, we express the solution in terms of the physical displacement i.e. $x = \sqrt{2/\omega_x} \operatorname{Re} (b_x) \equiv A_x \cos (\Omega t - \phi_x)$, and likewise for y-displacement. In order to extract the shape of the elliptical trajectory, we look at the covariance matrix of the oscillatory motion, which is:
\begin{equation}
\label{cov_matrix_single_pillar}
C = 
\begin{pmatrix}
\langle x^2 \rangle & \langle xy \rangle
\\
\langle xy \rangle & \langle y^2 \rangle
\end{pmatrix}
=
\frac{1}{2}
\begin{pmatrix}
A_x^2 & A_x A_y \cos (\phi_x - \phi_y)
\\
A_x A_y \cos (\phi_x - \phi_y) & A_y^2
\end{pmatrix}.
\end{equation}
We could then calculate eigenvalues and eigenvectors of this oscillator-motion covariance matrix, to obtain the parameters of the ellipse $A,\theta, \zeta$ (cf Figure~2
of the main text) as
\begin{align}
\label{ellipse_params}
A \cos \vert \zeta \vert, A \sin \vert \zeta \vert &= \frac{A_x A_y}{\sqrt{2}} \left[ \frac{1}{A_x^2} + \frac{1}{A_y^2} \pm \left( \left(\frac{1}{A_x^2} - \frac{1}{A_y^2} \right)^2 + \frac{4}{A_x^2 A_y^2}\cos ^2 (\phi_x - \phi_y) \right)^{1/2} \right] ^{1/2}, 
\\
\theta &= \arctan \left[ \frac{A_x A_y}{2\cos (\phi_x - \phi_y)} \left(\frac{1}{A_y^2} - \frac{1}{A_x^2} + \left( \left(\frac{1}{A_x^2} - \frac{1}{A_y^2} \right)^2 + \frac{4}{A_x^2 A_y^2}\cos ^2 (\phi_x - \phi_y) \right)^{1/2}  \right) \right], \\
\text{sign} (\zeta ) &= \text{sign} (xv_y - yv_x) = \text{sign} (\sin (\phi_y - \phi_x)).
\end{align}
In the latter equation, the sense of rotation of the ellipse is determined by the sign of the ellipticity $\zeta$, which can be determined from the sign of the angular momentum.
\section{Effective disorder model}\label{sec:disorder_theo}
\begin{figure}
\centering
\includegraphics[width=1\columnwidth]{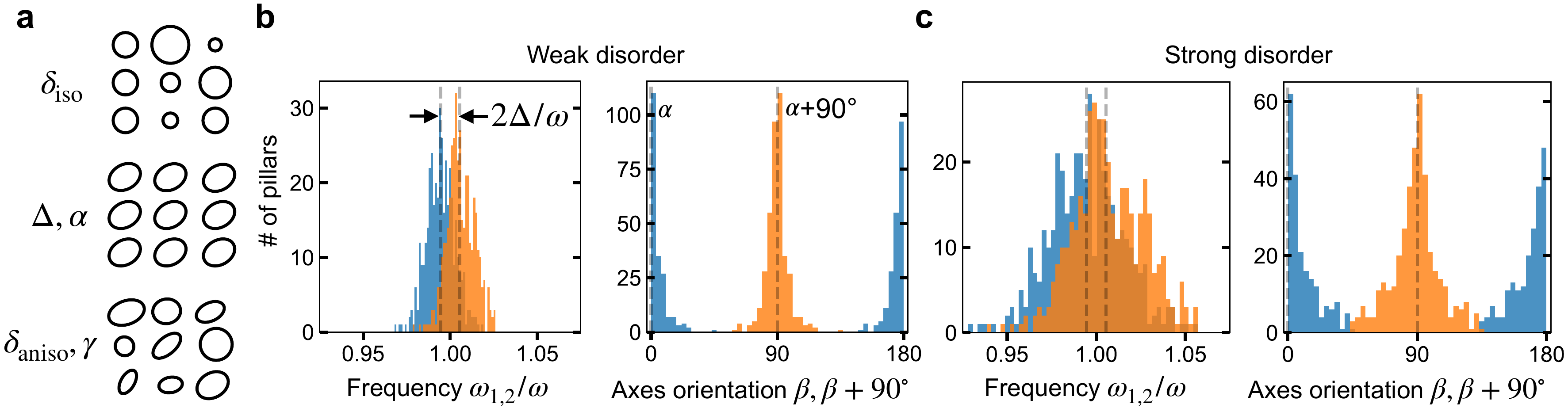}
\caption{{\bf Effective disorder model.} 
\textbf{a}, Schematic visualisation of the three different terms of the disordered on-site Hamiltonian (Equation~\ref{on_site_disordered_hamiltonian}). The 3x3 array of ellipses represent the different pillars in the array. The shape of the ellipses represents the top-view of the pillar in the experiment, whereas the size signifies the varying height of the pillars.
\textbf{b-c}, The distribution of uncoupled frequencies and orientations of all the pillars in the 20$\times$20 array for weak (\textbf{b}) and strong (\textbf{c}) disorder levels. The weak (strong) disorder case refers to the tight-binding parameters that are used in Fig.~5(\textbf{a}) (Figs.~3 and 4) of the main text. Tight-binding parameter values corresponding to the strong disorder is given in Appendix~\ref{sec:tight-binding_parameters}; for weak disorder, the parameters $\sigma_{\rm iso,aniso}$ are reduced to its 40\%.
}
\label{fig:SI_disorder_model}
\end{figure}
In the experimental array, the disorder primarily arises due to the variation of the pillar geometries in the fabrication process, which leads to a fluctuation of the isolated pillar frequencies. Further, we assume that the disorder in the coupling strengths between the neighboring pillars is negligible in comparison to the on-site disorder. Therefore, it can be ignored in the disorder model. In this section, we explicitly write down the disordered on-site Hamiltonian, and visualise the physical interpretation of the different tight-binding model parameters (see Figure~\ref{fig:SI_disorder_model}).

As explained in the main text, the mean parameters of an isolated pillar i.e. normal mode frequencies $\omega \pm \Delta$ and orientation $\alpha$ are perturbed in the presence of disorder. It is convenient to describe this perturbation in the matrix notation of the 2x2 disordered on-site Hamiltonian, as shown below
\begin{align}
\label{on_site_disordered_hamiltonian}
H_{\rm on-site}=
\begin{pmatrix}
    b_x^{*} & b_y^{*}
\end{pmatrix}
\biggl[
    \begin{pmatrix}
        \omega+\delta_{\rm iso} & 0 \\
        0 & \omega+\delta_{\rm iso}
    \end{pmatrix}
    +
    R(\alpha)
    \begin{pmatrix}
        -\Delta & 0 \\
        0 & \Delta
    \end{pmatrix}
    R^{-1}(\alpha)
    +
    R(\gamma)
    \begin{pmatrix}
        -\delta_{\rm aniso} & 0 \\
        0 & \delta_{\rm aniso}
    \end{pmatrix}
    R^{-1}(\gamma)
\biggr]
\begin{pmatrix}
    b_x \\
    b_y
\end{pmatrix}.
\end{align}
Here, $R(\alpha)$ is the rotation matrix given by Equation~\ref{rot_matrix}.

The physical interpretation of the three different terms of Equation~\ref{on_site_disordered_hamiltonian} is visualised in Figure~\ref{fig:SI_disorder_model}\textbf{a}. The disorder in the mean frequency $\omega$ (anisotropy parameter $\Delta$) is simulated by the random variable $\delta_{\rm iso} (\delta_{\rm aniso})$, that is sampled from a gaussian distribution with mean zero and standard deviation $\sigma_{\rm iso}$ $(\sigma_{\rm aniso})$. The fluctuation in the mean orientation $\alpha$ is characterised by the random angle parameter $\gamma$, which is sampled from a uniform distribution in the range $[0,2\pi)$. For a particular choice of disorder parameters, the distribution of the uncoupled frequencies $\omega_{1,2}$ and the orientations $\beta$ can be represented by a histogram plot as shown in Figure~\ref{fig:SI_disorder_model}\textbf{b,c}. Note that for reasons of simplicity, we consider that the isolated pillar parameters $\omega_{1},\omega_{2}$ and $\beta$ are uncorrelated among all the pillars in the array.

\section{Next-to-nearest neighbor coupling terms of the Hamiltonian}
\label{app:nnn coupling}
In this section, we show the explicit form of the next-to-nearest coupling term $H_d$ of the tight-binding Hamiltonian.

The next-to-nearest coupling Hamiltonian $H_d$ can be favourably written in terms of the complex amplitudes in the two diagonal directions $b_{\mathbf{r},\pm} = (b_{\mathbf{r},x} \pm b_{\mathbf{r},y})/\sqrt{2}$ ($\textbf{r}$ is the position of the pillar), as shown below
\begin{align}
\label{nnn hamiltonian}
H_d= -\underbrace{J_{\rm d,ll} \sum _{t,\langle \langle \mathbf{r},\mathbf{r}' \rangle \rangle_{t}} b^{*}_{\mathbf{r},t}b_{\mathbf{r}',t}}_{\text{n.n.n longitudinal coupling}} 
-
\underbrace{ J_{\rm d,tt} \sum _{t,\langle \langle \mathbf{r},\mathbf{r}' \rangle \rangle_{\bar{t}}} b^{*}_{\mathbf{r},t}b_{\mathbf{r}',t}}_{\text{n.n.n transversal coupling}}.
\end{align}
Here, $t=\{+,-\}$ labels the two diagonal directions, $\langle \langle \mathbf{r},\mathbf{r}' \rangle \rangle _t$ indicates the next-to-nearest neighbor in the $t$ direction. The bar symbol $\bar{t}$ in the transversal coupling term interchanges the two diagonal directions i.e. $\bar{+}=-$ and vice-versa.

\section{Approximate values of tight-binding parameters for the experimental system}\label{sec:tight-binding_parameters}
\renewcommand{\arraystretch}{1.25}
\begin{table}
    \centering
    \begin{tabular}{c|c|c}
        \hline
        \textbf{Parameter} & \textbf{Approximate value} & \textbf{Normalised value} \\
        \hline
        Mean frequency $\omega$ & $2\pi \times$ \SI{1.33\pm0.02}{\mega\hertz} &  1 \\
        Mean frequency anisotropy $\Delta$ & $2\pi \times$ \SI{7.5\pm2.5}{\kilo\hertz} &   0.006 \\
        Mean orientation of smaller frequency mode $\beta$ & \SI{0\pm3}{\degree} & $0^{\circ}$ \\
        Isotropic disorder $\sigma_{\rm iso}$ & $2\pi \times$ \num{27}$^{+22}_{-9}\,$\si{\kilo\hertz} & 0.020 \\
        Anisotropic disorder $\sigma_{\rm aniso}$ & $2\pi \times$ \SI{6.8\pm5}{\kilo\hertz} & 0.005  \\
        Longitudinal coupling (n.n) $J_{\rm ll}$ & $2\pi \times$ \SI{30\pm25}{\kilo\hertz} & 0.022 \\
        Longitudinal coupling (n.n.n) $J_{\rm d,ll}$ & $2\pi \times$ \SI{15\pm13}{\kilo\hertz} & 0.011 \\
        Transversal coupling (n.n) $J_{\rm tt}$ & $2\pi \times$ \SI{30\pm25}{\kilo\hertz} & 0.022 \\
        Transversal coupling (n.n.n) $J_{\rm d,tt}$ & $2\pi \times$ \SI{7.5\pm6}{\kilo\hertz} & 0.006 \\
        Mechanical damping $\Gamma$ & $2\pi \times$ \SI{5\pm1}{\kilo\hertz} &  0.004 \\
        \hline
    \end{tabular}
    \caption{\textbf{Estimated values of the tight-binding parameters for the experimental array in the main text.} The error in the approximate values (number in parentheses denotes the error such that the last digit of the error and quoted value correspond to each other) are estimated from either of the following: experimental data shown in the main text, the weakly coupled array in Figure~\ref{fig:SI_tight_binding_parameters}, the study of the coupling on pillar pairs in \cite{Doster.2019}, comparing the experimental and simulated response spectrum. The normalised frequency parameter values are evaluated relative to the mean frequency $\omega$.
    }
    \label{tab:estimated_parameters}
\end{table}
\begin{figure}
\centering
\includegraphics[width=1\columnwidth]{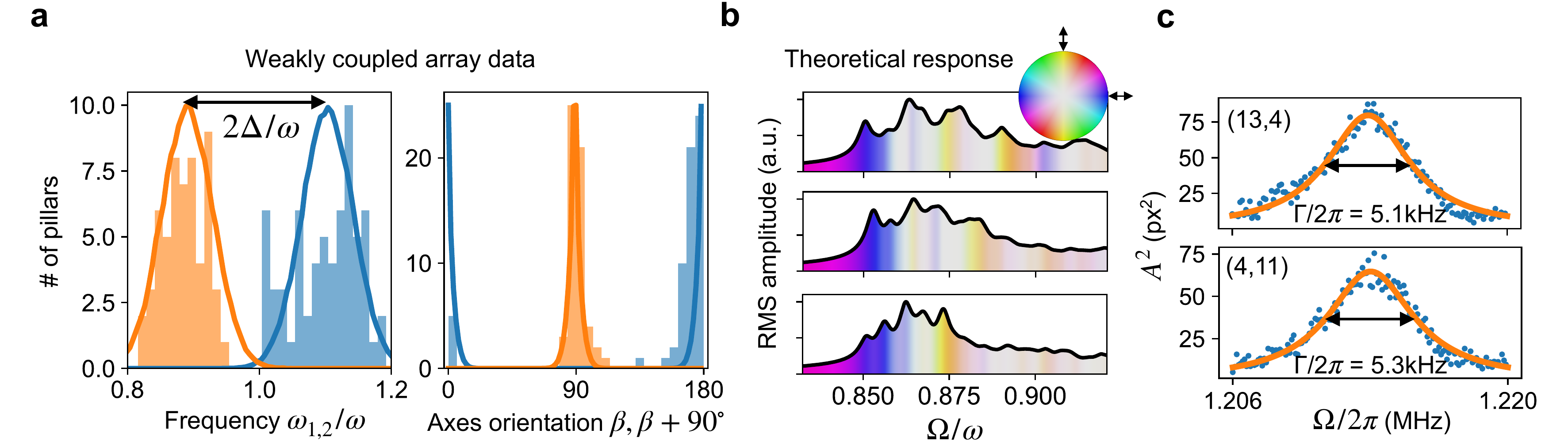}
\caption{{\bf Determining approximate values of tight-binding parameters.} 
\textbf{a}, Experimentally obtained distribution of isolated pillar frequencies and orientations of a very weakly coupled array. The isolated pillar parameters for this array are extracted by fitting this distribution (fit indicated with lines). 
\textbf{b}, Theoretically obtained response spectrums (with identical color scale as Figure~4 of the main text) for three different random realizations, but identical disorder parameters extracted from (a). 
\textbf{c}, Square of amplitude as a function of frequency for a single moving pillar mode. The FWHM of the fitted curve (orange) indicates the mechanical damping $\Gamma \approx 2\pi \times$ \SI{5}{\kilo\hertz}.
}
\label{fig:SI_tight_binding_parameters}
\end{figure}

As mentioned in the main text, it is impracticable to estimate the exact values of the tight-binding parameters. 
The major bottleneck behind estimating the isolated pillar parameters ($\omega, \Delta, \alpha, \delta_{\rm iso}, \delta_{\rm aniso}$) from the experimental data is that the trajectories are influenced due to the interaction with the neighbors. We came up with a solution to this conflict: to observe an array where the distance between pillars is sufficiently large that it is reasonable to ignore the couplings between the pillars. For such a very weakly coupled array, we reliably estimate the isolated pillar frequencies and orientations, and analyse its spatial profile (see Figure~\ref{fig:disorder_exp}) and statistics (see Figure~\ref{fig:SI_tight_binding_parameters}\textbf{a}). We use this weakly coupled array data as a guide to estimate the approximate values of the isolated pillar parameters for the array used in the main text, cf. Table~\ref{tab:estimated_parameters}.

Below, we present the approximate values of the 10 tight-binding parameters for the array presented in the main text.

\begin{itemize}
    \item Mean frequency $\omega$: It is roughly equal to the mean of the two strongest resonances in the response spectrum (RMS amplitude vs. frequency) in Figure~4 of the main text i.e. $\omega \approx 2\pi \times $\SI{1.35}{\mega\hertz}. 
    
    \item Mean frequency anisotropy $\Delta$: If it is too large $\Delta \gg \{\Gamma, J_{\rm ll}, J_{\rm tt}, J_{\rm d,ll}, J_{\rm d,tt}, \sigma_{\rm iso}, \sigma_{\rm aniso} \}$ (as is the case for the weakly coupled array), then the response spectrum would feature two different peaks corresponding to two orthogonal modes. If it is too small, then we would observe just a single band in the response spectrum. However, the experimentally observed response spectrum in the main text show none of the two cases, instead it exhibits resolved peaks with no clear sense of orientation within each peak. For $\Delta = 2\pi \times $ \SI{7.55}{\kilo\hertz} ($\Delta / \omega = 0.006$), we could qualitatively create this scenario in our theoretical model.
    
    \item Mean orientation of smaller frequency mode $\beta$: We have observed multiple arrays where frequency anisotropy $\Delta$ was large enough that we could associate orthogonal orientations to the two resonances in the response spectrum. For all such arrays, including the weakly coupled array data in Figure~\ref{fig:SI_tight_binding_parameters}\textbf{a}, the two normal-mode orientations are close to $0^\circ$ and $90^\circ$. For the experimental array in the main text, we observe that the trajectories are near horizontally polarized for smaller driving frequencies. Hence, we consider the smaller frequency mode to be at $0^\circ$.
    
    \item Disorder parameters $\sigma_{\rm iso}$ and $\sigma_{\rm aniso}$: 
    The theoretical model assumes that the disorder mainly arises in the isolated pillar parameters $\omega$ and $\Delta$. For the weakly coupled array, we extract the disorder parameters as: $\sigma_{\rm iso}/\omega = 0.036$ and $\sigma_{\rm aniso}/\Delta = 0.155$, cf. Figure~\ref{fig:SI_tight_binding_parameters}\textbf{a}. However, the isotropic disorder appears to be too strong for the experimental array in the main text. This is because the response spectrum for these parameters exhibits a rough lineshape (see Figure~\ref{fig:SI_tight_binding_parameters}\textbf{b}), in contrast to the smooth peaks in the main text. Therefore, we take the approximate isotropic disorder as $\sigma_{\rm iso}/\omega = 0.02$ (smaller than than that of uncoupled array). The diversity of the orientations in the two strongest resonances of the main text could be explained by selecting a larger anisotropic disorder $\sigma_{\rm aniso}/\Delta = 0.907$ than the weakly coupled array. The distribution of the estimated isolated pillar frequencies and orientations for the array in the main text is shown in Figure~\ref{fig:SI_disorder_model}\textbf{c}.
    
    \item Coupling parameters $J_{\rm ll}, J_{\rm d,ll}, J_{\rm tt}, J_{\rm d,tt}$: Using the experimental study of the interaction strength as a function of distance between the pillars \cite{Doster.2019}, we estimate that the coupling strengths are of order \SI{10}{\kilo\hertz}. We find that the theoretical model predicitions for $J_{\rm ll}=J_{\rm tt}=$ \SI{30}{\kilo\hertz} agrees well with the experimental results. The diagonal neighbor coupling strengths are estimated to be smaller, cf. Table~\ref{tab:estimated_parameters}, because of the increased distance between the pillars.
    
    \item Mechanical damping $\Gamma$: In the frequency response, we look at the cases when only one (or maximum two) pillar is effectively moving. By fitting the individual pillar response spectrum with a Lorentzian for such cases, we estimate the mechanical damping as $\Gamma \approx 2\pi \times $ \SI{5}{\kilo\hertz} (see Figure~\ref{fig:SI_tight_binding_parameters}\textbf{c}).
\end{itemize}

\section{Determination of the location of L lines and C points in the steady state pattern}
\label{app:topo singularities}
\begin{figure}
\centering
\includegraphics[width=1\columnwidth]{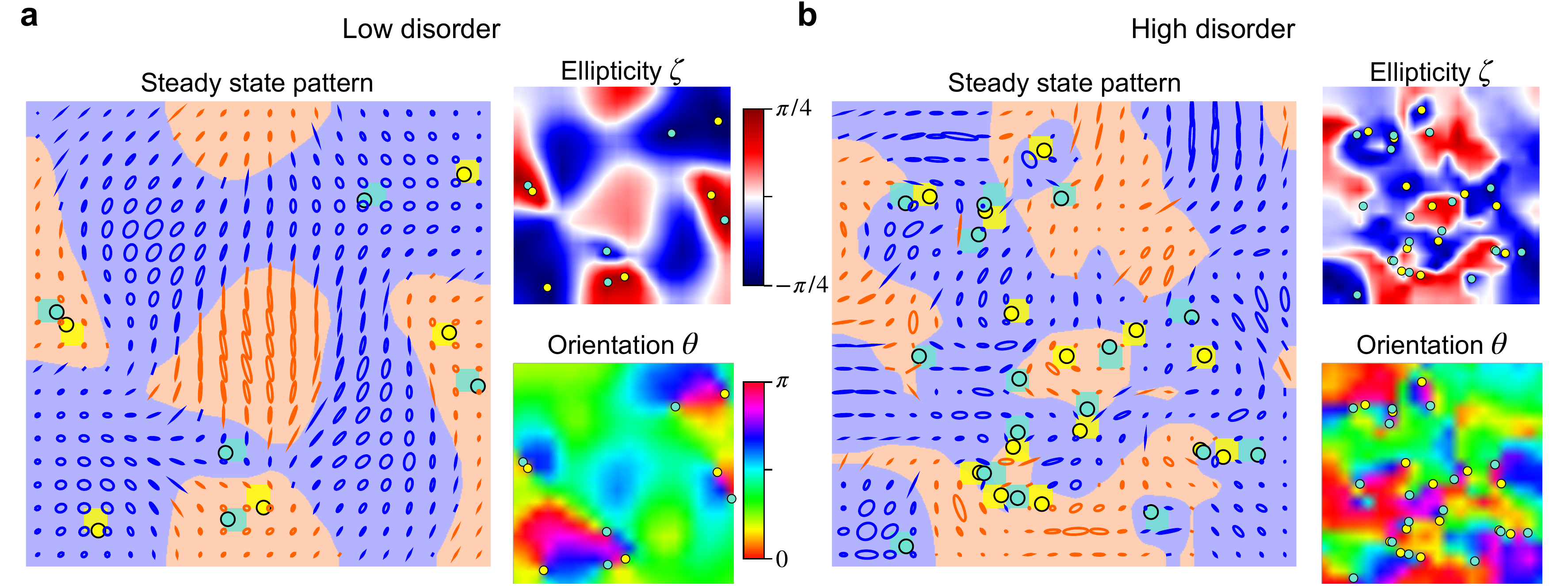}
\caption{{\bf Location of polarization singularities in theoretically obtained steady state pattern.} 
The position of L lines and C points are determined in the steady state patterns for weak (\textbf{a}) and strong (\textbf{b}) disorder levels. The ellipticity is zero at L lines, which separates region of opposite handedness (blue and red regions). At the C points, the major-axis orientation $\theta$ has a singularity with winding index $I=\pm 1/2$ (indicated with different marker colors) and $\zeta$ has its extremum. For stronger disorder in \textbf{b}, the field pattern fluctuates rapidly within the array. Therefore, there are several singularities in the polarization field, and the L lines and C points can easily be destroyed by slightly varying the drive frequency. Tight-binding parameter values corresponding to the weak (strong) disorder are similar as in Appendix~\ref{sec:tight-binding_parameters}, except the disorder parameters $\sigma_{\rm iso,aniso}$ are decreased (increased) by 60\% (100\%).
}
\label{fig:SI_topological_singularities}
\end{figure}

\begin{figure}
\centering
\includegraphics[width=1\columnwidth]{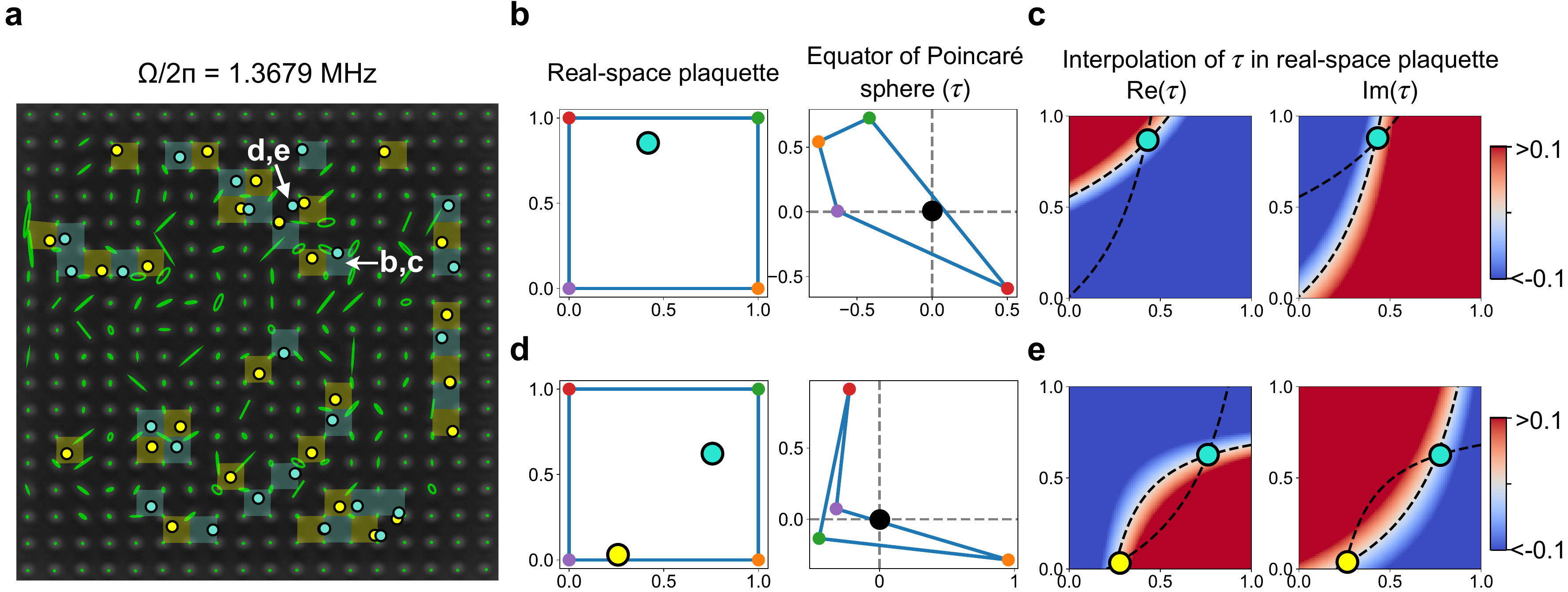}
\caption{\textbf{Determining sub-lattice resolved location of C points in the steady-state pattern.} 
\textbf{a}, Experimentally observed polarization pattern featuring C plaquettes ($I=\pm1/2$ for yellow (+) and blue (-)) and proposed location of C points within the plaquettes. 
\textbf{b-c} and \textbf{d-e}, Demonstration of the algorithm with two example plaquettes (indicated in \textbf{a}) containing different number of C points. 
\textbf{b,d}, The four elliptical trajectories at the plaquette corners are mapped on to the equatorial plane of the Poincar\'e sphere ($\tau$). The winding index of a plaquette is non-zero if the polygon constructed by joining the four mapped points encloses the C point (or the origin),  indicated as a black circle. 
\textbf{c,e}, The position of the C point in the real-space plaquette is obtained by finding the roots of the interpolated function $\tau (x,y)$ (zero contour-levels of Re($\tau$) and Im($\tau$) are indicated with dashed lines).
}
\label{fig:SI_interpolation}
\end{figure}
In Figure~5 
of the main text, we show the topologically robust L lines and C points in the steady state patterns. In this section, we describe the procedure of determining the locations of these polarization singularities, and convey that they are robust only for weak disorder levels.

In the first place, polarization singularities are defined for polarization fields in a continuous space, but here we consider the situation where the field is specified only on a discretized grid. However, the definition of polarization fields can be generalized also to this case in a consistent manner: If two neighboring grid points have opposite handedness ($\mathrm{sgn}\zeta$), they must be separated by an L line. If a plaquette has a non-trivial winding number for the orientation $\theta$, it must contain a C point. To determine smooth locations for the L lines and C points, we extend the field from the grid points to the entire plane by interpolation. The L lines can be obtained as the zero contours of the ellipticity $\zeta$, whereas C points are located at the nodes of the major-axis orientation $\theta$, cf. Figure~\ref{fig:SI_topological_singularities}\textbf{a}. 

More concretely, we determine the location of the C points within the array as follows: 

(i) First, we determine all the plaquettes where the C points are located. This can be done by considering the winding number of $\theta$ around a plaquette, which in the simplest case of a square lattice is
\begin{equation}
\label{winding_number}
I=\frac{d(\theta_{\rm sw},\theta_{\rm se}) + d(\theta_{\rm se},\theta_{\rm ne}) + d(\theta_{\rm ne},\theta_{\rm nw}) + d(\theta_{\rm nw},\theta_{\rm sw})}{2\pi},
\end{equation}
with
\begin{equation}
d(\theta_{\rm 1},\theta_{\rm 2}) = \theta_{\rm 2} - \theta_{\rm 1} + \pi \mathbb{Z} \in [-\pi /2,\pi /2).
\end{equation}
The indices stand for north/south-east/west, respectively. This quantity $I$ can be computed for a discrete grid, and its value being non-zero can be used to define a ``C plaquette''. Note that $I$ by definition can only be an integer multiple of $1/2$, since the differences of orientations $\theta_1-\theta_2$ cancel around the loop and only the "modulo $\pi$" operation in the definition of each individual term $d$ leads to a nonzero result. The sum of $I$ over all plaquettes in an infinitely extended lattice is conserved (in a finite system, changes may come in from the boundaries, just as for vortices). We also mention in passing that in principle, if the orientation field $\theta$ on the lattice is originally obtained by evaluation of an underlying smooth field (defined in continuous space), the plaquettes with  nonzero $I$ need not contain the locations of the $C$ points of this smooth field (they might e.g. sit in adjacent plaquettes, depending on the precise field configuration).

(ii) In principle, we could simply indicate the plaquettes with nonzero $I$. However, we can use interpolation to obtain a smooth version of the orientation field, which then enables us to propose a more precise location of the $C$ points (see Figure~\ref{fig:SI_interpolation}). In order to identify the location of the C point within a plaquette, we use the fact that C points are mapped to the poles of the Poincar\'e sphere (when considering the mapping from the real-space plaquette to this sphere). Therefore, they are vortices of the field 
$\tau = \cos (2\zeta) e^{2i\theta}$ (real and imaginary parts of $\tau$ represent the projection of a point onto the equatorial plane of the Poincar\'e sphere). 
Hence, for each plaquette, we are interested in the solution of
\begin{equation}
\tau(x_0+\delta x,y_0+\delta y) = 0,
\end{equation}
where $(x_0,y_0)$ is the bottom-left corner of this plaquette. Using bilinear interpolation, we determine the value of $\tau$ at any arbitrary point inside the plaquette, and numerically determine the roots of the above equation, cf. Figure~\ref{fig:SI_interpolation}. Note that it is possible to have multiple C points within the plaquette, subjecting to the condition that the sum of their winding index is equal to that of the winding index of the plaquette.

\begin{figure}
\centering
\includegraphics[width=1\columnwidth]{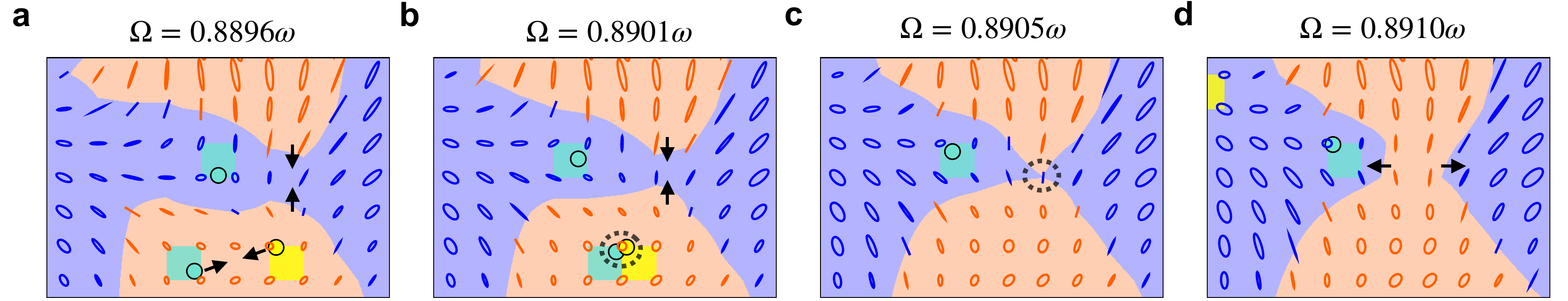}
\caption{\textbf{Robustness of topological singularities in the polarization patterns of slightly disordered arrays.} Simulation of the motion of topological singularities as a function of frequency. 
\textbf{a-d}, Frequency evolution of the steady state pattern in a section of a slightly disordered array. The L line (C point) is robust, as it can only deform and move around with frequency (black arrows), unless it merges with another another L line (C point of opposite winding index) and split (annihilate), see the black-dashed circle in \textbf{c} (\textbf{b}).
}
\label{fig:SI_topological_singularity_evolution}
\end{figure}

For weak disorder where the field is smooth, the polarization pattern in the vicinity of a C point is close to circular. One can then easily observe aspects such as how the annihilation  of two C points influences the surrounding polarization field (see Figure~\ref{fig:SI_topological_singularity_evolution}). For stronger disorder, the polarization pattern changes rapidly, with orientations of neighboring lattice points almost uncorrelated. 
Therefore, a larger fraction of the plaquettes are then classified as C plaquettes, and there are often pairs of directly neighboring plaquettes or even longer chains with C plaquettes of opposite winding index (see Figure~\ref{fig:SI_topological_singularities}\textbf{b}). They can easily disappear (e.g. when sweeping the frequency) because there is always a closeby annihilation partner. 
The stronger the disorder gets, the more this behaviour is observed.

\end{document}